\documentstyle[10pt,epsf,epsfig,hangcaption,xspace,amssymb,amsfonts,amsmath,amsthm,cite,dp_delphititle,lineno,rotating]{dp_delphi}

\makeindex
\def\DpPaperGroup{PH-EP}
\def\DpPaperRef{2004-036}
\def\DpDate{26 April 2004}
\def\DpAuthors{DELPHI Collaboration}
\def\DpSubmit{(Accepted by Euro. Phys. J.)}
\def\DpTitle{{  Measurement of the energy dependence of hadronic jet rates and
the strong coupling \boldmath\as\ from the four-jet rate with the DELPHI
detector at LEP }}
\def\DpComment{ }
\def\DpEMail{ }

\newcommand{\ecm}{E_{\mathrm cm}}

\newcommand{\fig}{Figure~\ref}
\newcommand{\tab}{Table~\ref}
\newcommand{\as}{$\alpha_s$}

\newcommand{\oas}{$\cal O$($\alpha_s^2$)}
\newcommand{\oass}{$\cal O$($\alpha_s^3$)}
\newcommand{\oasss}{$\cal O$($\alpha_s^4$)}
\newcommand{\gev}{\mbox{\,Ge\kern-0.15exV}}
\newcommand{\GeV}{\mbox{\,Ge\kern-0.15exV}}
\newcommand{\mev}{\mbox{\,Me\kern-0.15exV}}

\newcommand{\qcd}{$e^+e^- \rightarrow Z/\gamma \rightarrow
q\bar{q}$\hspace{0.1cm}}

\newcommand{\beq}{\begin{equation}}
\newcommand{\eeq}{\end{equation}}
\def\jetset{{JETSET}}
\def\pythia{{PYTHIA}}

\def\herwig{{HERWIG}}
\def\ariadne{{ARIADNE}}
\def\apacic{{APACIC\kern-0.15ex+\kern-0.15ex+}}
\def\amegic{{Amegic\kern-0.15ex+\kern-0.15ex+}}
\def\sprime{{SPRIME}}
\def\sprimep{{SPRIME\kern-0.15ex+}}
\def\delsim{{DELSIM}}
\def\zfitter{{Zfitter}}

\def\delphi{{DELPHI}}

\def\lep{{LEP}}
\def\lepone{{LEP}1}
\def\leptwo{{LEP}2}

\newcommand{\vjr} {Vierjetrate}

\newcommand{\cambridge} {{CAMBRIDGE}}

\newcommand{\durham} {{DURHAM}}
\newcommand{\jade} {{JADE}}

\newcommand{\eqn}{Eq.~\ref}

\newcommand{\sect}{Sec.~\ref}
\newcommand{\asmz}{$\alpha_s(M_Z^2)$}

\newcommand{\xmu}{$x_{\mu}$}
\newcommand{\xmuopt}{$x_{\mu}^{\mathrm{opt}}$}
\newcommand{\ycut}{$y_{\mathrm{cut}}$}

\newcommand{\debrecen}{{DEBRECEN}}
\newcommand{\excalibur}{{EXCALIBUR}}

\newcommand{\cpp}{{\sc C\kern-0.25ex+\kern-0.15ex+}}

\newcommand{\myhangcaption}[2]{ \begin{sc} \hangcaption[#1]{\sl#2} \end{sc} } 

\begin{document}
\makeatletter
\makeatother

\begin{titlepage}
\pagenumbering{roman}

\CERNpreprint{\DpPaperGroup}{\DpPaperRef} 
\date{{\small\DpDate}} 
\title{\DpTitle} 
\address{\DpAuthors} 

\begin{shortabs} 
\noindent
\noindent
Hadronic events from the data collected with the \delphi\ detector
at \lep\ within the energy range from 89\gev\ to 209\gev\ are
selected, their jet rates are determined and compared to
predictions of four different event generators. One of them is the
recently developed APACIC++ generator which
performs a massive matrix element calculation matched to a parton
shower followed by string fragmentation. The four-jet rate is used
to measure $\alpha_s$ in the next-to-leading-order approximation
yielding
\begin{eqnarray}
  \mbox{\asmz} = 0.1175 \pm 0.0030. \nonumber
\end{eqnarray}
The running of $\alpha_s$ determined by using four-jet events has
been tested. The logarithmic energy slope is measured to be
\begin{eqnarray}
 \frac{{\mathrm d}\alpha_s^{-1}}{{\mathrm d} \log E_{\mathrm {cm}}}
         =1.14 \pm 0.36.  \nonumber
\end{eqnarray}
Since the analysis is based on four-jet final states it represents
an alternative approach to previous \delphi\ $\alpha_s$
measurements using event shape distributions.

\end{shortabs}

\vfill

\begin{center}
\DpSubmit \ \\ 
\DpComment \ \\
\DpEMail \ \\
\end{center}

\vfill
\clearpage

\headsep 10.0pt

\addtolength{\textheight}{10mm}
\addtolength{\footskip}{-5mm}
\begingroup
%
\newcommand{\DpName}[2]{\hbox{#1$^{\ref{#2}}$},\hfill}
\newcommand{\DpNameTwo}[3]{\hbox{#1$^{\ref{#2},\ref{#3}}$},\hfill}
\newcommand{\DpNameThree}[4]{\hbox{#1$^{\ref{#2},\ref{#3},\ref{#4}}$},\hfill}
\newskip\Bigfill \Bigfill = 0pt plus 1000fill
\newcommand{\DpNameLast}[2]{\hbox{#1$^{\ref{#2}}$}\hspace{\Bigfill}}

%
\footnotesize
\noindent
\DpName{J.Abdallah}{LPNHE}
\DpName{P.Abreu}{LIP}
\DpName{W.Adam}{VIENNA}
\DpName{P.Adzic}{DEMOKRITOS}
\DpName{T.Albrecht}{KARLSRUHE}
\DpName{T.Alderweireld}{AIM}
\DpName{R.Alemany-Fernandez}{CERN}
\DpName{T.Allmendinger}{KARLSRUHE}
\DpName{P.P.Allport}{LIVERPOOL}
\DpName{U.Amaldi}{MILANO2}
% ...
\normalsize
\endgroup

\titlefoot{Department of Physics and Astronomy, Iowa State
     University, Ames IA 50011-3160, USA
    \label{AMES}}
\titlefoot{Physics Department, Universiteit Antwerpen,
     Universiteitsplein 1, B-2610 Antwerpen, Belgium \\
     \indent~~and IIHE, ULB-VUB,
     Pleinlaan 2, B-1050 Brussels, Belgium \\
     \indent~~and Facult\'e des Sciences,
     Univ. de l'Etat Mons, Av. Maistriau 19, B-7000 Mons, Belgium
    \label{AIM}}
\titlefoot{Physics Laboratory, University of Athens, Solonos Str.
     104, GR-10680 Athens, Greece
    \label{ATHENS}}
\titlefoot{Department of Physics, University of Bergen,
     All\'egaten 55, NO-5007 Bergen, Norway
    \label{BERGEN}}
\titlefoot{Dipartimento di Fisica, Universit\`a di Bologna and INFN,
     Via Irnerio 46, IT-40126 Bologna, Italy
    \label{BOLOGNA}}
\titlefoot{Centro Brasileiro de Pesquisas F\'{\i}sicas, rua Xavier Sigaud 150,
     BR-22290 Rio de Janeiro, Brazil \\
     \indent~~and Depto. de F\'{\i}sica, Pont. Univ. Cat\'olica,
     C.P. 38071 BR-22453 Rio de Janeiro, Brazil \\
     \indent~~and Inst. de F\'{\i}sica, Univ. Estadual do Rio de Janeiro,
     rua S\~{a}o Francisco Xavier 524, Rio de Janeiro, Brazil
    \label{BRASIL}}
\titlefoot{Coll\`ege de France, Lab. de Physique Corpusculaire, IN2P3-CNRS,
     FR-75231 Paris Cedex 05, France
    \label{CDF}}
\titlefoot{CERN, CH-1211 Geneva 23, Switzerland
    \label{CERN}}
\titlefoot{Institut de Recherches Subatomiques, IN2P3 - CNRS/ULP - BP20,
     FR-67037 Strasbourg Cedex, France
    \label{CRN}}
\titlefoot{Now at DESY-Zeuthen, Platanenallee 6, D-15735 Zeuthen, Germany
    \label{DESY}}
\titlefoot{Institute of Nuclear Physics, N.C.S.R. Demokritos,
     P.O. Box 60228, GR-15310 Athens, Greece
    \label{DEMOKRITOS}}
\titlefoot{FZU, Inst. of Phys. of the C.A.S. High Energy Physics Division,
     Na Slovance 2, CZ-180 40, Praha 8, Czech Republic
    \label{FZU}}
\titlefoot{Dipartimento di Fisica, Universit\`a di Genova and INFN,
     Via Dodecaneso 33, IT-16146 Genova, Italy
    \label{GENOVA}}
\titlefoot{Institut des Sciences Nucl\'eaires, IN2P3-CNRS, Universit\'e
     de Grenoble 1, FR-38026 Grenoble Cedex, France
    \label{GRENOBLE}}
\titlefoot{Helsinki Institute of Physics, P.O. Box 64,
     FIN-00014 University of Helsinki, Finland
    \label{HELSINKI}}
\titlefoot{Joint Institute for Nuclear Research, Dubna, Head Post
     Office, P.O. Box 79, RU-101 000 Moscow, Russian Federation
    \label{JINR}}
\titlefoot{Institut f\"ur Experimentelle Kernphysik,
     Universit\"at Karlsruhe, Postfach 6980, DE-76128 Karlsruhe,
     Germany
    \label{KARLSRUHE}}
\titlefoot{Institute of Nuclear Physics,Ul. Kawiory 26a,
     PL-30055 Krakow, Poland
    \label{KRAKOW1}}
\titlefoot{Faculty of Physics and Nuclear Techniques, University of Mining
     and Metallurgy, PL-30055 Krakow, Poland
    \label{KRAKOW2}}
\titlefoot{Universit\'e de Paris-Sud, Lab. de l'Acc\'el\'erateur
     Lin\'eaire, IN2P3-CNRS, B\^{a}t. 200, FR-91405 Orsay Cedex, France
    \label{LAL}}
\titlefoot{School of Physics and Chemistry, University of Lancaster,
     Lancaster LA1 4YB, UK
    \label{LANCASTER}}
\titlefoot{LIP, IST, FCUL - Av. Elias Garcia, 14-$1^{o}$,
     PT-1000 Lisboa Codex, Portugal
    \label{LIP}}
\titlefoot{Department of Physics, University of Liverpool, P.O.
     Box 147, Liverpool L69 3BX, UK
    \label{LIVERPOOL}}
\titlefoot{Dept. of Physics and Astronomy, Kelvin Building,
     University of Glasgow, Glasgow G12 8QQ
    \label{GLASGOW}}
\titlefoot{LPNHE, IN2P3-CNRS, Univ.~Paris VI et VII, Tour 33 (RdC),
     4 place Jussieu, FR-75252 Paris Cedex 05, France
    \label{LPNHE}}
\titlefoot{Department of Physics, University of Lund,
     S\"olvegatan 14, SE-223 63 Lund, Sweden
    \label{LUND}}
\titlefoot{Universit\'e Claude Bernard de Lyon, IPNL, IN2P3-CNRS,
     FR-69622 Villeurbanne Cedex, France
    \label{LYON}}
\titlefoot{Dipartimento di Fisica, Universit\`a di Milano and INFN-MILANO,
     Via Celoria 16, IT-20133 Milan, Italy
    \label{MILANO}}
\titlefoot{Dipartimento di Fisica, Univ. di Milano-Bicocca and
     INFN-MILANO, Piazza della Scienza 2, IT-20126 Milan, Italy
    \label{MILANO2}}
\titlefoot{IPNP of MFF, Charles Univ., Areal MFF,
     V Holesovickach 2, CZ-180 00, Praha 8, Czech Republic
    \label{NC}}
\titlefoot{NIKHEF, Postbus 41882, NL-1009 DB
     Amsterdam, The Netherlands
    \label{NIKHEF}}
\titlefoot{National Technical University, Physics Department,
     Zografou Campus, GR-15773 Athens, Greece
    \label{NTU-ATHENS}}
\titlefoot{Physics Department, University of Oslo, Blindern,
     NO-0316 Oslo, Norway
    \label{OSLO}}
\titlefoot{Dpto. Fisica, Univ. Oviedo, Avda. Calvo Sotelo
     s/n, ES-33007 Oviedo, Spain
    \label{OVIEDO}}
\titlefoot{Department of Physics, University of Oxford,
     Keble Road, Oxford OX1 3RH, UK
    \label{OXFORD}}
\titlefoot{Dipartimento di Fisica, Universit\`a di Padova and
     INFN, Via Marzolo 8, IT-35131 Padua, Italy
    \label{PADOVA}}
\titlefoot{Rutherford Appleton Laboratory, Chilton, Didcot
     OX11 OQX, UK
    \label{RAL}}
\titlefoot{Dipartimento di Fisica, Universit\`a di Roma II and
     INFN, Tor Vergata, IT-00173 Rome, Italy
    \label{ROMA2}}
\titlefoot{Dipartimento di Fisica, Universit\`a di Roma III and
     INFN, Via della Vasca Navale 84, IT-00146 Rome, Italy
    \label{ROMA3}}
\titlefoot{DAPNIA/Service de Physique des Particules,
     CEA-Saclay, FR-91191 Gif-sur-Yvette Cedex, France
    \label{SACLAY}}
\titlefoot{Instituto de Fisica de Cantabria (CSIC-UC), Avda.
     los Castros s/n, ES-39006 Santander, Spain
    \label{SANTANDER}}
\titlefoot{Inst. for High Energy Physics, Serpukov
     P.O. Box 35, Protvino, (Moscow Region), Russian Federation
    \label{SERPUKHOV}}
\titlefoot{J. Stefan Institute, Jamova 39, SI-1000 Ljubljana, Slovenia
     and Laboratory for Astroparticle Physics,\\
     \indent~~Nova Gorica Polytechnic, Kostanjeviska 16a, SI-5000 Nova Gorica, Slovenia, \\
     \indent~~and Department of Physics, University of Ljubljana,
     SI-1000 Ljubljana, Slovenia
    \label{SLOVENIJA}}
\titlefoot{Fysikum, Stockholm University,
     Box 6730, SE-113 85 Stockholm, Sweden
    \label{STOCKHOLM}}
\titlefoot{Dipartimento di Fisica Sperimentale, Universit\`a di
     Torino and INFN, Via P. Giuria 1, IT-10125 Turin, Italy
    \label{TORINO}}
\titlefoot{INFN,Sezione di Torino, and Dipartimento di Fisica Teorica,
     Universit\`a di Torino, Via P. Giuria 1,\\
     \indent~~IT-10125 Turin, Italy
    \label{TORINOTH}}
\titlefoot{Dipartimento di Fisica, Universit\`a di Trieste and
     INFN, Via A. Valerio 2, IT-34127 Trieste, Italy \\
     \indent~~and Istituto di Fisica, Universit\`a di Udine,
     IT-33100 Udine, Italy
    \label{TU}}
\titlefoot{Univ. Federal do Rio de Janeiro, C.P. 68528
     Cidade Univ., Ilha do Fund\~ao
     BR-21945-970 Rio de Janeiro, Brazil
    \label{UFRJ}}
\titlefoot{Department of Radiation Sciences, University of
     Uppsala, P.O. Box 535, SE-751 21 Uppsala, Sweden
    \label{UPPSALA}}
\titlefoot{IFIC, Valencia-CSIC, and D.F.A.M.N., U. de Valencia,
     Avda. Dr. Moliner 50, ES-46100 Burjassot (Valencia), Spain
    \label{VALENCIA}}
\titlefoot{Institut f\"ur Hochenergiephysik, \"Osterr. Akad.
     d. Wissensch., Nikolsdorfergasse 18, AT-1050 Vienna, Austria
    \label{VIENNA}}
\titlefoot{Inst. Nuclear Studies and University of Warsaw, Ul.
     Hoza 69, PL-00681 Warsaw, Poland
    \label{WARSZAWA}}
\titlefoot{Fachbereich Physik, University of Wuppertal, Postfach
     100 127, DE-42097 Wuppertal, Germany
    \label{WUPPERTAL}}
\addtolength{\textheight}{-10mm}
\addtolength{\footskip}{5mm}
\clearpage

\headsep 30.0pt
\end{titlepage}

%
\pagenumbering{arabic} 
\setcounter{footnote}{0} %
\large
%
\section{Introduction}
%
Measurements of hadronic multijet rates in electron-positron
annihilation provide an excellent test 
of perturbative quantum chromodynamics (QCD). They can be 
confronted with predictions of QCD-based hadronisation models and
allow a precise determination of the strong coupling \as.
Furthermore, the study of multijet production originating from QCD
processes is essential for the understanding of the background to
four-quark production in $W^+W^-$ or $ZZ$ decays and also for
understanding the background in the search for new phenomena.
Here we report the final measurements of 2-, 3-, 4-, and 5-jet
rates using all data collected by  \delphi\ 
during the years 1993 to 2000. The 4-jet rate is used to determine
\as. 

Until 1995 the large electron-positron storage ring \lep\ at
CERN operated at centre-of-mass energies around the $Z$ resonance.
Due to the high cross-section the total number of hadronic events
collected during this part of the \lepone\ phase is about 2.5
million. Analysing \lepone\ data enables precise measurements of
the strong coupling and detailed comparisons of different methods
for extracting \as, see e.g. \cite{Abreu:2000ck}. From autumn
1995 onwards the centre-of-mass energy was continuously increased
and finally reached about 209 GeV in October 2000. During the
\leptwo\ programme DELPHI collected a total of about 12000
hadronic $q\bar{q}$ events at centre-of-mass energies between 130
GeV and 209 GeV.
The statistics of hadronic events collected at \leptwo, though
small compared to that gathered near the $Z$ resonance,
are sufficient for the measurement of jet rates and for a
determination of the strong coupling \as, see e.g.
\cite{Abreu:1999rc}. Analysing both \lepone\ and \leptwo\ data
gives access to the energy dependence, the running of the strong
coupling and thus to
a direct test of asymptotic freedom.  

In \sect{sec_select} the selection of hadronic events, the
reconstruction of the centre-of-mass energy, the correction
procedures applied to the data and the suppression of $W^+W^-$ and
$ZZ$ events (and other four-fermion background) are briefly
discussed. \sect{results} presents the applied jet clustering
algorithms, the measured jet rates and the comparison of the data
with predictions from hadronic event generators. In \sect{alphas},
the measurement of \as\ based on the 4-jet data is presented. As
in all analyses using topological information from hadronic
events, the error on the value of \as\ is dominated by theoretical
uncertainties.  
Here we
determine  \as\ by applying second order perturbation theory with
an optimised renormalisation scale. In \sect{alphas}, the \as\
measurements along with studies of different choices of the
renormalisation scale and the investigation of the running of \as\
from \lepone\ and \leptwo\ data are presented.

\section{Selection and correction of hadronic data\label{sec_select}}
The analysis uses data taken with the \delphi\ detector at 
centre-of-mass energies between 89\gev\ and 209\gev\ divided into 14
energy bins. Data entering the analysis, including the integrated
luminosities collected at these energies and the cross-sections of
the contributing processes, are summarised in \tab{tab:energien}.

\begin{table}[tbhp]
  \renewcommand{\arraystretch}{1.3}
  \begin{center}
    \begin{tabular}{|c|r@{.}l|c|r@{.}l|r@{.}l|r@{.}l|r@{.}l|r@{.}l|r|}
      \cline{2-15}
        \multicolumn{1}{c|}{}&\multicolumn{2}{|c|}{}&&\multicolumn{2}{|c|}{}&\multicolumn{2}{|c|}{}&\multicolumn{2}{|c|}{}&\multicolumn{2}{|c|}{}&\multicolumn{2}{|c|}{}&\\  
        \multicolumn{1}{c|}{}&\multicolumn{2}{|c|}{}&&\multicolumn{2}{|c|}{}&\multicolumn{2}{|c|}{}&\multicolumn{2}{|c|}{}&\multicolumn{2}{|c|}{}&\multicolumn{2}{|c|}{}&\\
        \multicolumn{1}{c|}{}&\multicolumn{2}{|c|}{}&&\multicolumn{2}{|c|}{}&\multicolumn{2}{|c|}{}&\multicolumn{2}{|c|}{}&\multicolumn{2}{|c|}{}&\multicolumn{2}{|c|}{}&\\
        \multicolumn{1}{c|}{}&\multicolumn{2}{|c|}{}&&\multicolumn{2}{|c|}{}&\multicolumn{2}{|c|}{}&\multicolumn{2}{|c|}{}&\multicolumn{2}{|c|}{}&\multicolumn{2}{|c|}{}&\\
        \multicolumn{1}{c|}{}&
        \multicolumn{2}{|c|}{
          \hspace{0.25cm}\begin{rotate}{90}
          $E_{\mathrm{cm}} \left[ \mathrm{GeV} \right]$
          \end{rotate}
        } &
        \hspace{0.25cm}\begin{rotate}{90}year\end{rotate} &
        \multicolumn{2}{|c|}{\hspace{0.25cm}\begin{rotate}{90}${\cal L}\left[ \mathrm{pb}^{-1} \right]$\end{rotate}} &
        \multicolumn{2}{|c|}{\hspace{0.25cm}\begin{rotate}{90}$\sigma_{q\bar q}\left[\mathrm{pb} \right]$\end{rotate}} &
        \multicolumn{2}{|c|}{\hspace{0.25cm}\begin{rotate}{90}$\sigma_{q\bar q}^{\sqrt{s^\prime_{\mathrm {rec}}} > 0.9 \cdot \sqrt{s} }\left[\mathrm{pb} \right]$\end{rotate}} & 
        \multicolumn{2}{|c|}{\hspace{0.25cm}\begin{rotate}{90}$\sigma_{{\tiny{WW}}} \left[ \mathrm{pb} \right]$\end{rotate}} & 
        \multicolumn{2}{|c|}{\hspace{0.25cm}\begin{rotate}{90}$\sigma_{{\tiny{ZZ}}} \left[ \mathrm{pb} \right]$\end{rotate}} & 
        \multicolumn{1}{|c|}{\hspace{0.25cm}\begin{rotate}{90}$N_{\mathrm{hadr.}}$\end{rotate}} \\ 
      \cline{2-15} 
      \multicolumn{15}{c}{\vspace{-0.58cm}} \\ 
      \hline 
        & 
        89&4 & 
        1993/95 & 
        18&6 & 
        9900&  & 
        \multicolumn{2}{|c|}{---}  & 
        \multicolumn{2}{|c|}{---}  & 
        \multicolumn{2}{|c|}{---}  & 
        163\,013 \\ 
      \cline{2-15} 
        & 
        91&2 & 
        1993/94/95 & 
        77&4 & 
        30\,400&  & 
        \multicolumn{2}{|c|}{---} & 
        \multicolumn{2}{|c|}{---} & 
        \multicolumn{2}{|c|}{---} & 
        2\,091\,448 \\ 
      \cline{2-15} 
        \hspace{0.25cm}\begin{rotate}{90}\,\, \lepone \end{rotate} 
        & 
        93&0 & 
        1993/95 & 
        13&8 & 
        14100&  & 
        \multicolumn{2}{|c|}{---} & 
        \multicolumn{2}{|c|}{---} & 
        \multicolumn{2}{|c|}{---} & 
        237\,674 \\ 
      \hline 
      \hline 
        & 
        \multicolumn{2}{|c|}{ } & 
        1995 & 
        \multicolumn{2}{|c|}{ } & 
        \multicolumn{2}{|c|}{ } & 
        \multicolumn{2}{|c|}{ } & 
        \multicolumn{2}{|c|}{ } & 
        \multicolumn{2}{|c|}{ } & \\ 
      \cline{4-4} 
        & 
        \multicolumn{2}{|r|}{\raisebox{2.0ex}[-2.0ex]{133.2}} & 
        1997 & 
        \multicolumn{2}{|r|}{\raisebox{2.0ex}[-2.0ex]{11.9}} & 
        \multicolumn{2}{|r|}{\raisebox{2.0ex}[-2.0ex]{292.0}}  & 
        \multicolumn{2}{|r|}{\raisebox{2.0ex}[-2.0ex]{69.2}} & 
        \multicolumn{2}{|r|}{\raisebox{2.0ex}[-2.0ex]{---}}  & 
        \multicolumn{2}{|r|}{\raisebox{2.0ex}[-2.0ex]{---}}  & 
        \raisebox{2.0ex}[-2.0ex]{846} \\ 
      \cline{2-15} 
        & 
        161&4 & 
        & 
        11&5 & 
        147&0  & 
        32&3 & 
        3&4  & 
        \multicolumn{2}{|c|}{---} & 
        358 \\ 
      \cline{2-3} \cline{5-15} 
        & 
        172&3 & \raisebox{2.0ex}[-2.0ex]{1996} & 
        10&8 & 
        121&0  & 
        27&5 & 
        12&3 & 
        \multicolumn{2}{|c|}{---} & 
        261 \\ 
      \cline{2-15} 
        & 
        183&1 & 
        1997 & 
        57&9 & 
        100&3  & 
        23&4 & 
        16&5 & 
        1&0 & 
        1\,173 \\ 
      \cline{2-15} 
        & 
        189&2 & 
        1998 & 
        157&0 & 
        99&8  & 
        21&1 & 
        17&5 & 
        1&6 & 
        3\,053 \\ 
      \cline{2-15} 
        & 
        192&2 & 
        & 
        25&2 & 
        96&0  & 
        20&2 & 
        18&1 & 
        1&7 & 
        466 \\ 
      \cline{2-3} \cline{5-15} 
        \hspace{0.25cm}\begin{rotate}{90}\,\,\,\,\,\,\,\,\, \leptwo\ \end{rotate} & 
        196&2 & 
        & 
        78&4 & 
        90&0  & 
        19&2 & 
        18&6 & 
        1&7 & 
        1\,338 \\ 
      \cline{2-3} \cline{5-15} 
        & 
        200&1 & 
        \raisebox{2.0ex}[-2.0ex]{1999} & 
        81&8 & 
        85&2  & 
        18&2 & 
        18&7 & 
        1&8 & 
        1\,339 \\ 
      \cline{2-3} \cline{5-15} 
        & 
        202&1 & 
        & 
        39&8 & 
        83&3  & 
        17&7 & 
        18&8 & 
        1&8 & 
        642 \\ 
      \cline{2-15} 
        & 
        204&9 & 
        & 
        76&1 & 
        80&0  & 
        17&0 & 
        18&9 & 
        1&8 & 
        1\,187 \\ 
      \cline{2-3} \cline{5-15} 
        & 
        206&8 & 
        \raisebox{2.0ex}[-2.0ex]{2000} & 
        84&1 & 
        77&7  & 
        16&5 & 
        18&9 & 
        1&8 & 
        1\,297 \\ 
      \hline 
    \end{tabular} 
    \myhangcaption{analysed data} 
                {\label{tab:energien}Data entering the analysis: the columns 
                show the mean centre-of-mass energies $\ecm$, the years of
                data  
                taking, the 
                integrated luminosities, the cross-sections for $q\bar{q}$
                (before and  
                after the cut on the effective centre-of-mass energy
                $\sqrt{s^{\prime}_{\mathrm{rec}}} >0.9\cdot
                E_{\mathrm{cm}}$, described in the text),  
                $W^+W^-$, and neutral boson pair production 
                (from \zfitter 6.21 
                \cite{Bardin:1999yd}) and the number of selected $q\bar{q}$ 
                events after the cuts described in the text.} 
  \end{center} 
  \renewcommand{\arraystretch}{1.0} 
\end{table} 
%
%
%
%
%
%
 
\delphi\ is a hermetic detector with a solenoidal magnetic field 
of 1.2T. The tracking detectors in the barrel part (starting from 
the beam pipe) are a silicon micro-vertex detector (VD), a combined 
jet/proportional chamber inner detector (ID), a time projection 
chamber (TPC) as the main tracking device, and a streamer tube 
detector (OD) in the barrel region. The forward region is covered by 
silicon mini-strip and pixel detectors (VFT) and by the drift 
chamber detectors (FCA and FCB). 
 
The electromagnetic calorimeters are the high density projection 
chamber (HPC) in the barrel, and the lead-glass calorimeter (FEMC) in 
the forward region. The hadron calorimeter (HCAL) is a sampling gas detector
incorporated in the magnet yoke. Detailed information about the design and 
performance of \delphi\ can be found 
in~\cite{Aarnio:1991vx,Abreu:1996uz}.

In order to select well measured charged particle tracks, 
the cuts given in the upper part of \tab{cuts} have been applied. 
The cuts in the lower part of the table are used to 
select \qcd\ events and to 
suppress background processes such as two-photon interactions, beam-gas and 
beam-wall interactions, leptonic final states, and, for the \leptwo\ analysis, 
initial state radiation (ISR) and four-fermion (4f) background. 

\newcommand{\bmincut}{\leq 0.08} 
\newcommand{\sprimecut}{\geq  0.9 \cdot \ecm} 
\begin{table}[bt] 
\begin{center} 
\renewcommand{\arraystretch}{1.2} 
\begin{tabular}{|l|l|l|}\hline 
 Track    & $0.4\gev\leq p \leq 100\gev$ \\ 
 selection   & $ \Delta p / p \leq 1.0$     \\ 
             & measured track length $\geq 30$ cm \\ 
             & distance to I.P in $r\phi$ plane $\leq 4$ cm \\ 
             & distance to I.P. in $z \leq 10$ cm  \\ 
 
\hline 
\hline 
 
 Event       & $  N_{\mathrm {charged}}\ge 7 $ \\ 
 selection   & $ 25^{\circ} \le \theta_{\mathrm {thrust}} \le 155^{\circ} $ 
               \\ 
               \hline 
 ISR cuts    & $ E_{\mathrm {tot}} \geq  0.50\cdot\ecm $  \\ 
             & $ \sqrt{s^\prime_{\mathrm {rec}}} \sprimecut $ \\ 
             \hline 
 $WW$ and 4f cuts     &   $D^{2} > 900\gev^2$    \\ 
             & $ 42\ge N_{\mathrm {charged}} $ \\ 
             \hline 
\end{tabular} 
\myhangcaption{gaga}{\label{cuts}%
Criteria for track- and event selection. $p$ is the momentum, 
$\Delta p$ its error, $r$  the radial distance to the beam-axis, 
$z$  the distance to the beam interaction point (I.P.) along the 
beam-axis, $\phi$  the azimuthal angle, $N_{\mathrm  {charged}}$ 
the number of charged particles, $\theta_{\mathrm {thrust}}$ the 
polar angle of the thrust axis with respect to the beam, 
$E_{\mathrm {tot}}$ the total energy carried by all measured particles, 
$\sqrt{s'_{\mathrm {rec}}}$ the effective centre-of-mass 
energy, $\ecm=2 E_{beam}=\sqrt{s}$ the nominal centre-of-mass 
energy, and $D^2$ the discrimination variable, defined in 
\eqn{eq:defd2}. The first two cuts apply to charged and neutral 
particles, while the other track selection cuts apply only to 
charged particles. } \vspace{-0.7cm} 
\end{center} 
\end{table} 

At energies above 91.2\gev, the large cross-section of the $Z$ 
resonance peak raises the possibility of hard intial state radiation (ISR) 
allowing the 
creation of a nearly on-shell $Z$ boson. These ``radiative return 
events'' constitute a large fraction of all hadronic events. The 
ISR photons are typically  aligned along the beam 
direction and usually escape detection. In order 
to evaluate the effective hadronic centre-of-mass energy 
$\sqrt{s^{\prime}}$ of an event, considering ISR, an algorithm 
called \sprime\ is used~\cite{ABREU:1998yx}. \sprime\ is based on 
a fit imposing four-momentum conservation to measured jet 
four-momenta (including estimates of their errors). The hypotheses 
of single and multi photon radiation are tested based on the 
$\chi^2$ obtained in the corresponding constrained fits.
 
\begin{figure}[tb] 
\unitlength1cm 
 \unitlength1cm 
 \begin{center} 
 \begin{minipage}[t]{7.5cm} 
           \mbox{\epsfig{file=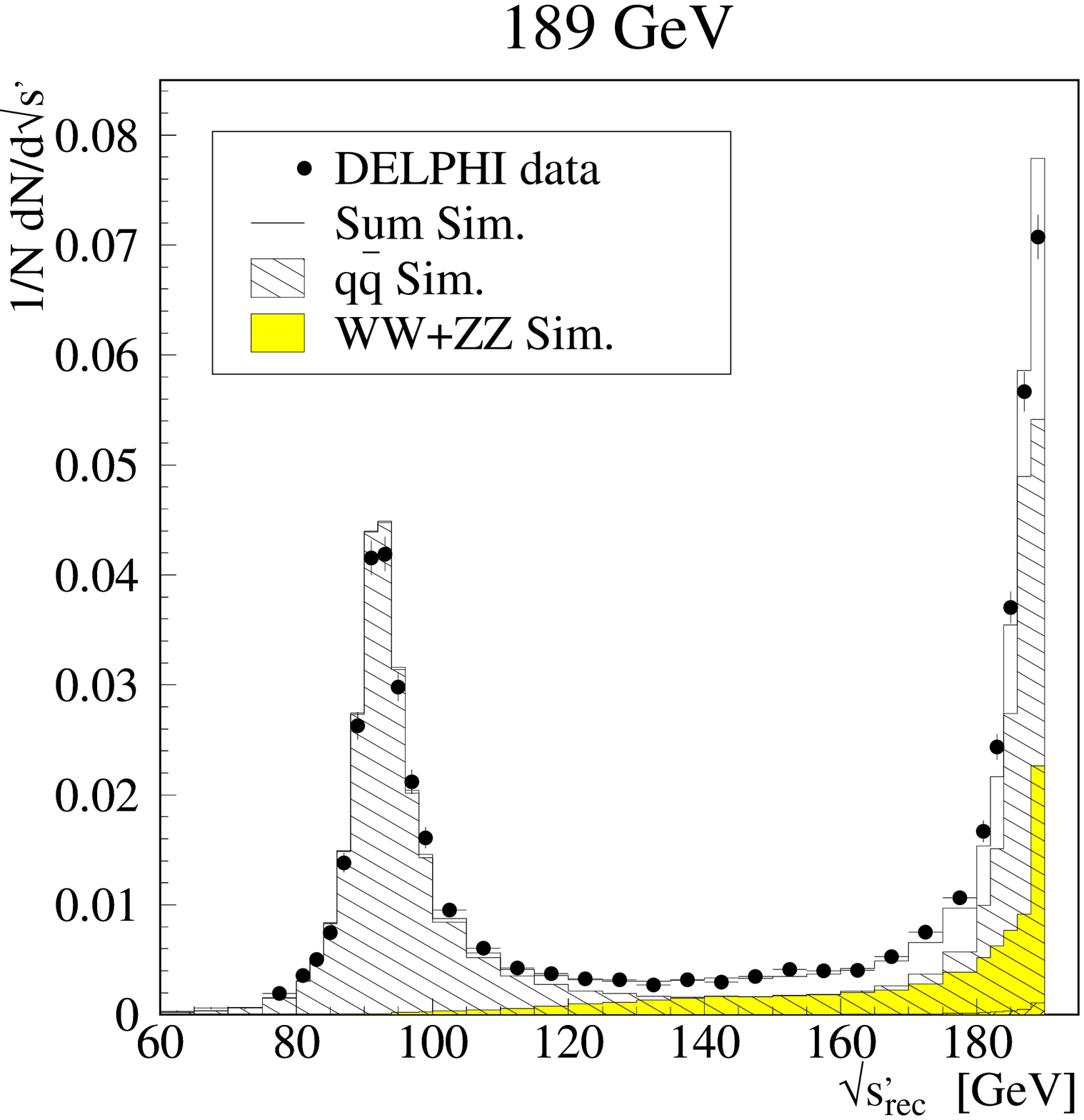,width=8.0cm}} 
 \end{minipage} 
 \hspace{.3cm} 
 \begin{minipage}[t]{7.5cm} 
           \mbox{\epsfig{file=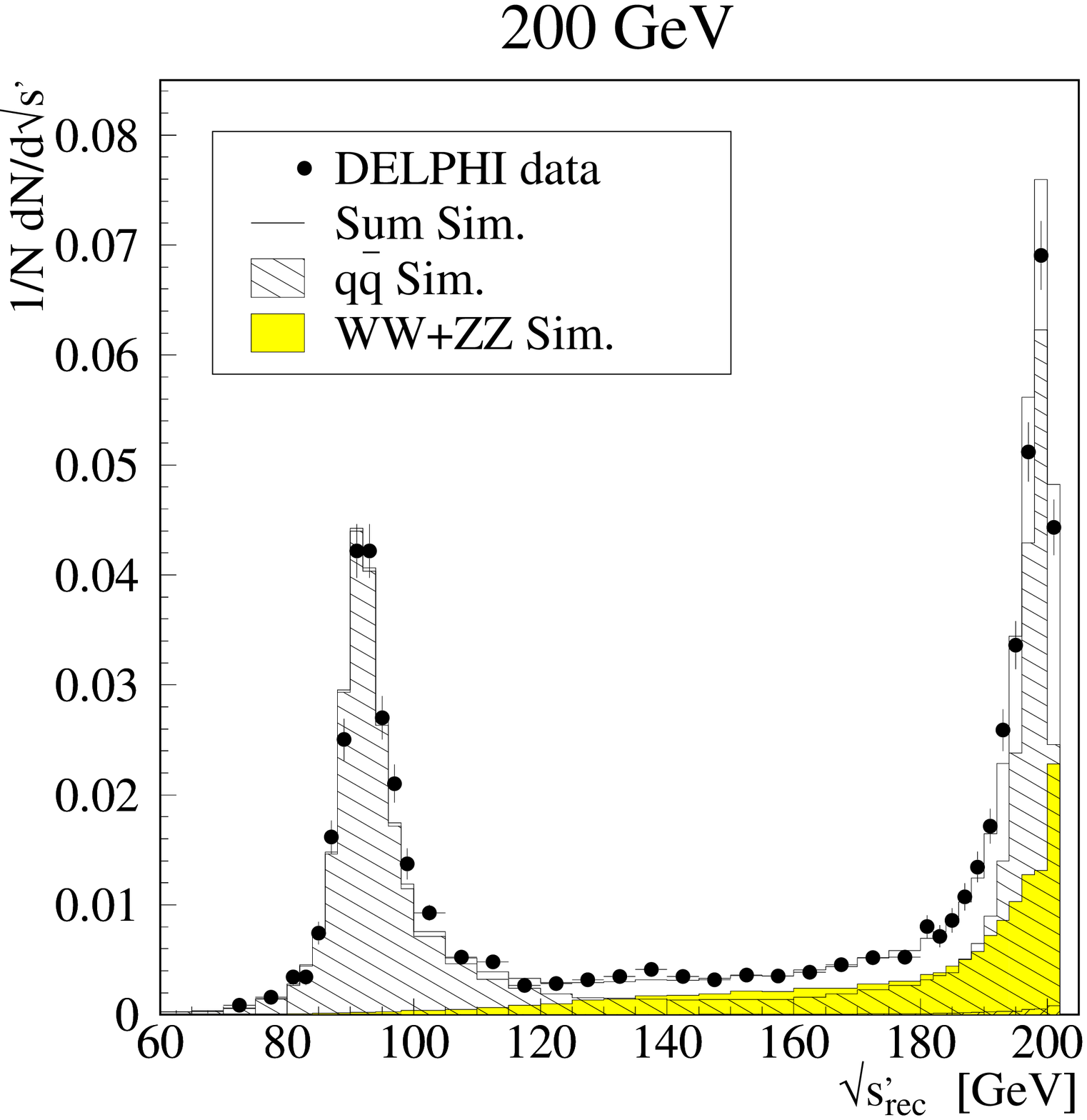,width=8.0cm}} 
 \end{minipage} 
\end{center} 
\myhangcaption{SPRIME}{\label{plot_isr}Reconstructed centre of 
  mass energy $\sqrt{s^{\prime}_{\mathrm {rec}}}$ before all cuts except the
  one on $\sqrt{s^{\prime}_{\mathrm rec}}$ 
  compared to QCD and four-fermion simulations. The small differently shaded
  areas in the bottom right of the plots indicate the size of WW and neutral
  boson pair background. 
} 
\end{figure}

\fig{plot_isr} shows the 189 and 200 \gev\, effective centre-of-mass energy
spectra as computed with \sprime\ for simulated and measured  events passing
all but the  
$\sqrt{s^{\prime}_{\mathrm {rec}}}$ cut. 
A cut on the reconstructed centre-of-mass energy 
$\sqrt{s'_{\mathrm {rec}}}\sprimecut$ is applied to discard 
radiative return events (see \tab{cuts}). Two-photon events, 
leptonic events and events due to leptonic or semileptonic 
$W^+W^-$ decays are strongly suppressed by the cuts. The remaining 
background from these types of events was found to be negligible 
in the following analysis. 
\begin{figure}[tb] 
\unitlength1cm 
 \unitlength1cm 
 \begin{center} 
 \begin{minipage}[t]{7.5cm} 
           \mbox{\epsfig{file=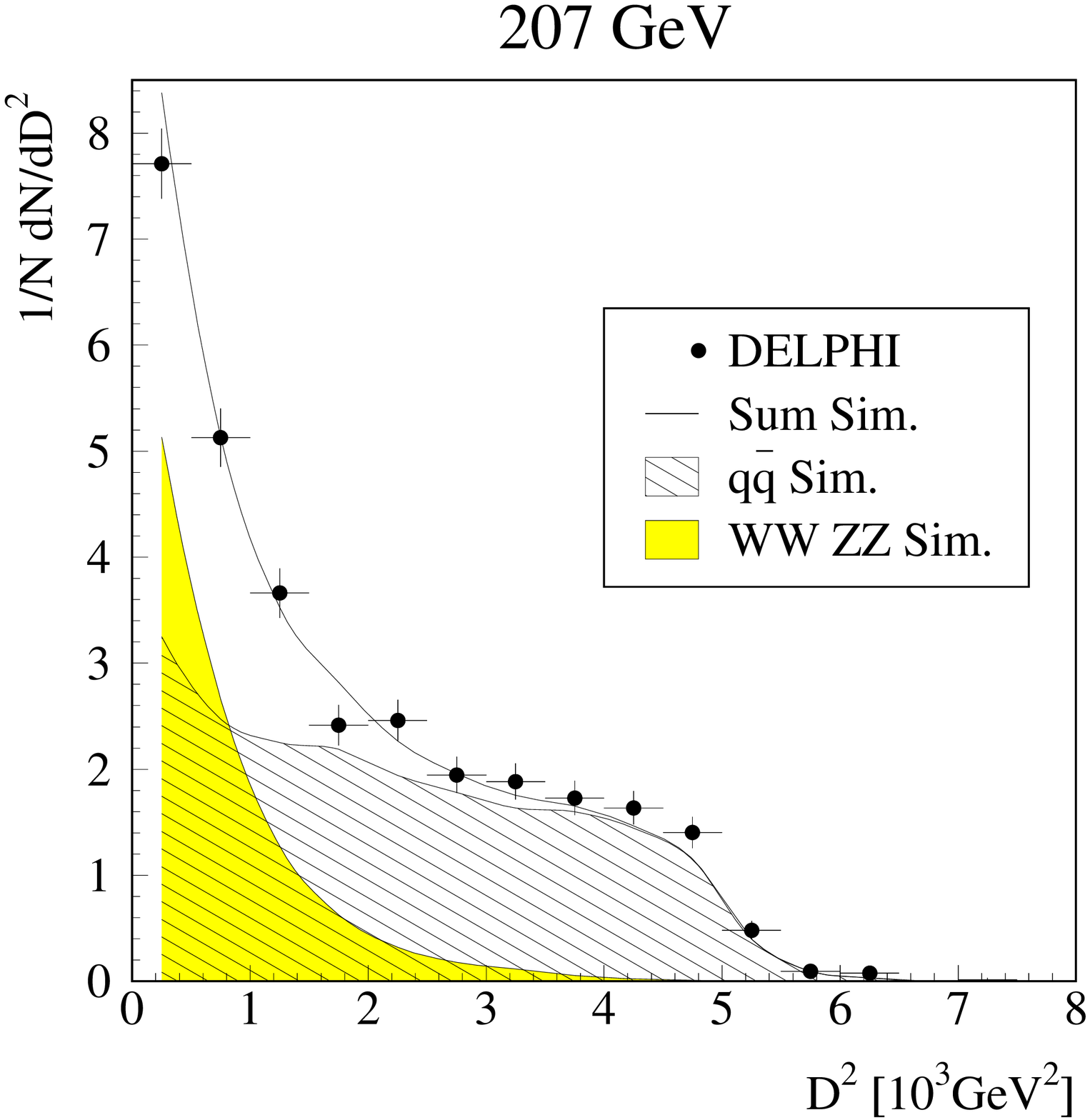,width=10.0cm}} 
 \end{minipage} 
\end{center} 
\myhangcaption{$D^2$}{\label{fig:d2}Distribution of $D^2$ for accepted data
at 207 GeV 
  before applying four-fermion cuts, compared to $q\bar{q}$ and 4f simulation.} 
\end{figure} 

Since the topological signatures of QCD four-jet events and hadronic 
4f events are similar, no highly 
efficient separation of the two classes of events is possible. 
Furthermore any 4f rejection implies a bias to the shape 
distributions of QCD events, which needs to be corrected with 
simulation. 
In this analysis a cut in the discrimination variable $D^2$
\cite{Becks:1999fa} is  applied to separate four-jet QCD events   
from hadronic $W^+W^-$ decays. All events are forced into a four-jet
configuration by a clustering algorithm. From the resulting four-momenta of 
the pseudo-particles the following quantity is calculated: 
%
\begin{eqnarray} 
     D^2 &=& \min{\left\{ (M_{ij}-M_W)^2 + 
                           (M_{kl}-M_W)^2 \right\} } \label{eq:defd2} \\ 
         &  & (ij;kl) = (12;34),(13;24),(14;23) \, .  \nonumber 
\end{eqnarray} 

The discrimination variable $D^2$ is based on a comparison of 
invariant dijet masses to the nominal mass of the $W$ boson. The 
minimum difference for all possible jet pairings $(ij,kl)$ is 
expected to be small for events arising from boson pair 
production. \fig{fig:d2} shows the distribution of $D^2$ at 205 and  
207\gev, compared to the simulation of contributing processes. 
Events from $W^+W^-$ or neutral boson pair production cluster at small 
values of $D^2$, while \qcd\ events extend to larger values of $D^2$.
Demanding $D^2 
> 900\gev^2$ leads to an efficient suppression of $W^+W^-$ and neutral boson
pair  
events. All remaining 4f contributions are estimated by using 
Monte Carlo generators and subtracted from the measurement. The 
simulations are normalised using the cross-sections given in 
\tab{tab:energien}. The quoted $\sigma_{WW}$ values 
correspond to a $W$ mass of 80.35\gev. For the simulation of WW and ZZ events
the following generators were used:
\begin{itemize}
\item    
\excalibur\ \cite{Berends:1995xn} generates four-fermion final
states through all possible electroweak four-fermion processes. 
The generator includes the width of the $W$ and $Z$  bosons. QED initial 
state corrections are implemented using a structure function formalism
\cite{Berends:1994}. EXCALIBUR also includes a Coulomb correction \cite{coul}
for the CC03 $WW$ production \cite{cc03}.   
\item 
\pythia~5.7~\cite{Sjostrand:1995iq} is a general-purpose Monte Carlo
generator for multi-particle production in high energy physics. As a 
general-purpose generator it  does not contain the detailed modelling
of all  the specific corrections that are contained in the dedicated
four-fermion generators. 
\end{itemize}
For the central result the \excalibur\ generator was applied while the
difference between \pythia\ and \excalibur\ was used to estimate the
systematic uncertainty on the background subtraction (see Sec.~\ref{run}).
 
Detector and cut effects are unfolded with simulation. The 
influence of detector effects  was studied by passing generated 
events (\jetset/\pythia~\cite{Sjostrand:1985ys} using the \delphi\ 
tuning described in~\cite{Abreu:1996na}) through a full detector 
simulation (\delsim~\cite{Aarnio:1991vx}). These Monte Carlo 
events are processed with the reconstruction program applying 
selection cuts as for the real data. In order to correct for cuts, 
detector and ISR effects, a bin-by-bin acceptance correction $C$, 
obtained from \qcd\ simulation, is applied to the data: 
 \begin{equation} 
  C_{det,i} = \frac{h(f_i)_{\mathrm {gen, no ISR}}} {h(f_i)_{\mathrm {acc}}}
  ,
\label{acc} 
 \end{equation} 
where $h(f_i)_{\mathrm {gen, no ISR}}$ represents 
 bin $i$ of the shape distribution $f$ 
generated with the tuned generator. The subscript $\mathrm {no 
ISR}$ indicates that only events without large ISR 
{($\sqrt{s}-\sqrt{s^\prime_{\mathrm {rec}}}<0.1\gev$)} enter the 
distribution. $h(f_i)_{\mathrm {acc}}$ represents the accepted 
distribution $f$ as obtained with the full detector simulation. 
%

 
\section{Jet rates} 
\label{results} 
Jet clustering algorithms are applied to cluster the large number 
of particles of a hadronic event into a small number of jets, 
reflecting the structure of hard partons of the event. Most clustering 
algorithms in $e^+ e^-$ annihilation apply a recursive scheme 
based on an ordering variable $d_{ij}$, a distance measure 
$y_{ij}$ and a merging scheme indicated by $\oplus$ in the 
following, all being functions of the four-momenta $p$ of two 
objects $i$ and $j$. The algorithms start with a table of 
particles representing the initial objects. The pair of objects 
with the smallest $d_{ij}$ is considered for merging. These two objects
are merged into one new object by applying the merging scheme $p_k = p_i 
\oplus p_j$, provided that the distance measure $y_{ij}$ is 
smaller than some given maximum separation $y_{\mathrm{cut}}$. 
This step is repeated with the two particles $i$ and $j$ replaced by
the combined object $k$. After each iteration the ordering variables 
$d_{ij}$ have to be recalculated.  The iteration stops if only one object
remains or if all distance measures $y_{ij}$ are larger than
$y_{\mathrm{cut}}$.

The remaining objects are called {\it jets} and the number of jets $n$ is
a function of the cutoff parameter $y_{\mathrm{cut}}$.

The $n$-jet rate, $R_n(y_{\mathrm cut})$ gives the fraction of $n$-jet events 
relative to all events. By definition:
\begin{equation}
  \sum_{i} R_i(y_{\mathrm cut}) \equiv 1 \, .
\end{equation}

The details of the clustering algorithms used in this analysis are defined
below.




\subsection{\jade} 
%
The \jade\ algorithm \cite{jade} is based on the same distance measure and 
ordering variable: 
\begin{equation} 
  d_{ij} = y_{ij} = \frac{2 E_i E_j \cdot \left(1 - \cos \theta_{ij}\right)} 
                {E_{\mathrm{vis}}^2} \, , 
  \label{jade-yij} 
\end{equation} 
$E_{\mathrm{vis}}$ being the visible energy, which would be the centre of 
mass energy $E_{\mathrm{cm}}$ for a perfect detector, $E_i,E_j$ being the 
energy of the objects $i$ and $j$ and $\theta_{ij}$ being the angle between 
$\vec{p}_i$ and $\vec{p}_j$. 
 
The merging scheme simply adds the four momenta of $p_i$ and 
$p_j$: 
\begin{equation} 
 p_k=p_i\oplus p_j =  p_i+p_j \label{jade_plus} \, . 
\end{equation} 
 
There are shortcomings within \jade\ arising from the choice of 
the distance measure $y_{ij}$. For events with soft gluons 
radiated off the quark and the antiquark, there are kinematical 
regions where \jade\ combines the soft gluons first. The resulting 
``phantom'' jet has a resultant momentum at large angle to the 
initial quarks and may point to a region where no particles exist.

\subsection{\durham} 
%
In case of the \durham\ or $k_{\bot}$ algorithm \cite{durham} the 
distance measure $d_{ij}$ and the 
ordering variable $y_{ij}$ are the same but they are now changed 
from mass to normalised transverse momentum $k_{\bot}$. 
\begin{equation} 
  d_{ij} = y_{ij} = \frac{2 \cdot \min\left\{E_i^2,E_j^2\right\} \cdot 
                 \left(1 - \cos \theta_{ij}\right)} 
                {E_{\mathrm{vis}}^2} \, . 
  \label{durham-yij} 
\end{equation} 
This choice mitigates the shortcomings of the \jade\ algorithm. 
 
\subsection{\cambridge} 
%
The \cambridge\ algorithm \cite{cambridge} is a modified $k_{\bot}$ clustering 
algorithm similar to the \durham\ algorithm. It is designed to 
preserve the advantages of \durham\ while reducing non-perturbative
corrections at small $y$ and providing better resolution of jet  
substructure. \cambridge\ is based on 
the same distance measure $y_{ij}$ as \durham\ 
({Eq.\ref{jade_plus}, \ref{durham-yij}}). 
The ordering variable $d_{ij}$ is a function of the angle between the 
objects $i$ and $j$ (``angular ordering''): 
\begin{equation} 
  d_{ij} = 2 \cdot \left(1 - \cos \theta_{ij}\right) \, . 
  \label{cambridge-order} 
\end{equation} 
 
If $y_{ij} \geq y_{\mathrm{cut}}$ the object with the smaller 
energy is stored as a jet and deleted from the event table (``soft 
freezing''). If $y_{ij} < y_{\mathrm{cut}}$ the objects are merged 
into a new object. The iteration stops if only one 
object remains or if all distance measures $y_{ij}$ are 
larger than $y_{\mathrm{cut}}$. 
 
\subsection{Results} 
\begin{figure}[tbh] 
  \begin{center} 
 \mbox{ \epsfig{file=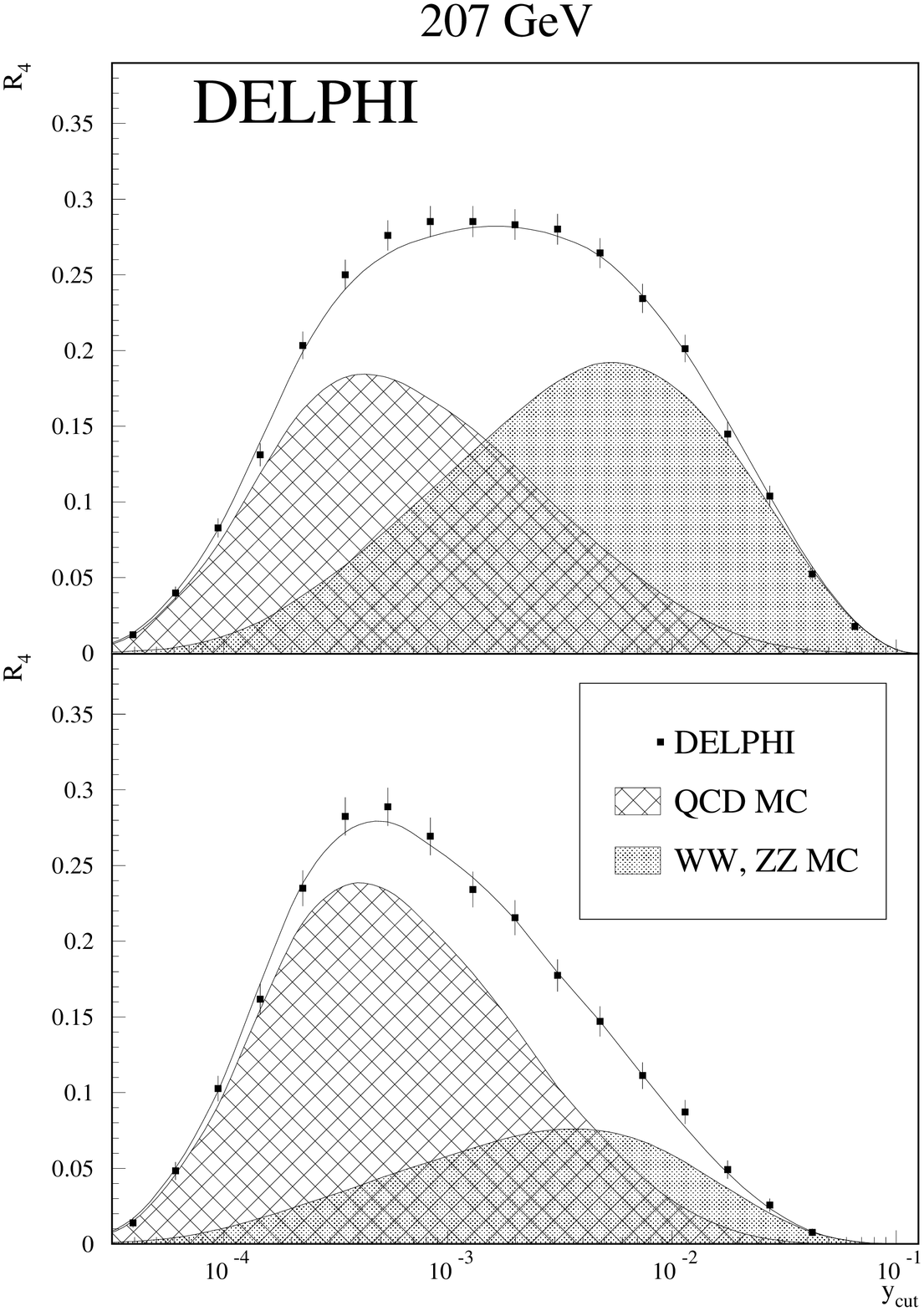,height=15cm} } 
  \end{center} 
  \vspace{0cm} 
  \myhangcaption{R4RAW}{\label{fig:raw_rates_207gev}Four-jet rate ($R_4$), 
  determined with the \cambridge\ algorithm, from raw data 
  at 207\gev\ as a function of \ycut, compared to the simulation of the 
  contributing processes. Top: before cuts against four-fermion 
  background. Bottom: After cutting at  $D^2>900\gev^2$.} 
\end{figure}

\fig{fig:raw_rates_207gev} shows the CAMBRIDGE four-jet rate $R_4$ as a 
function of \ycut\ from the 207\gev\ data before and after the 
$D^2>900\gev^2$ cut and underlines the separation power of $D^2$. 
The data are found to be in good agreement with the simulation.

\begin{figure}[tbhp] 
  \begin{center} 
   \vspace{-2cm} 
   \hspace{0cm} 
    \mbox{ \epsfig{file=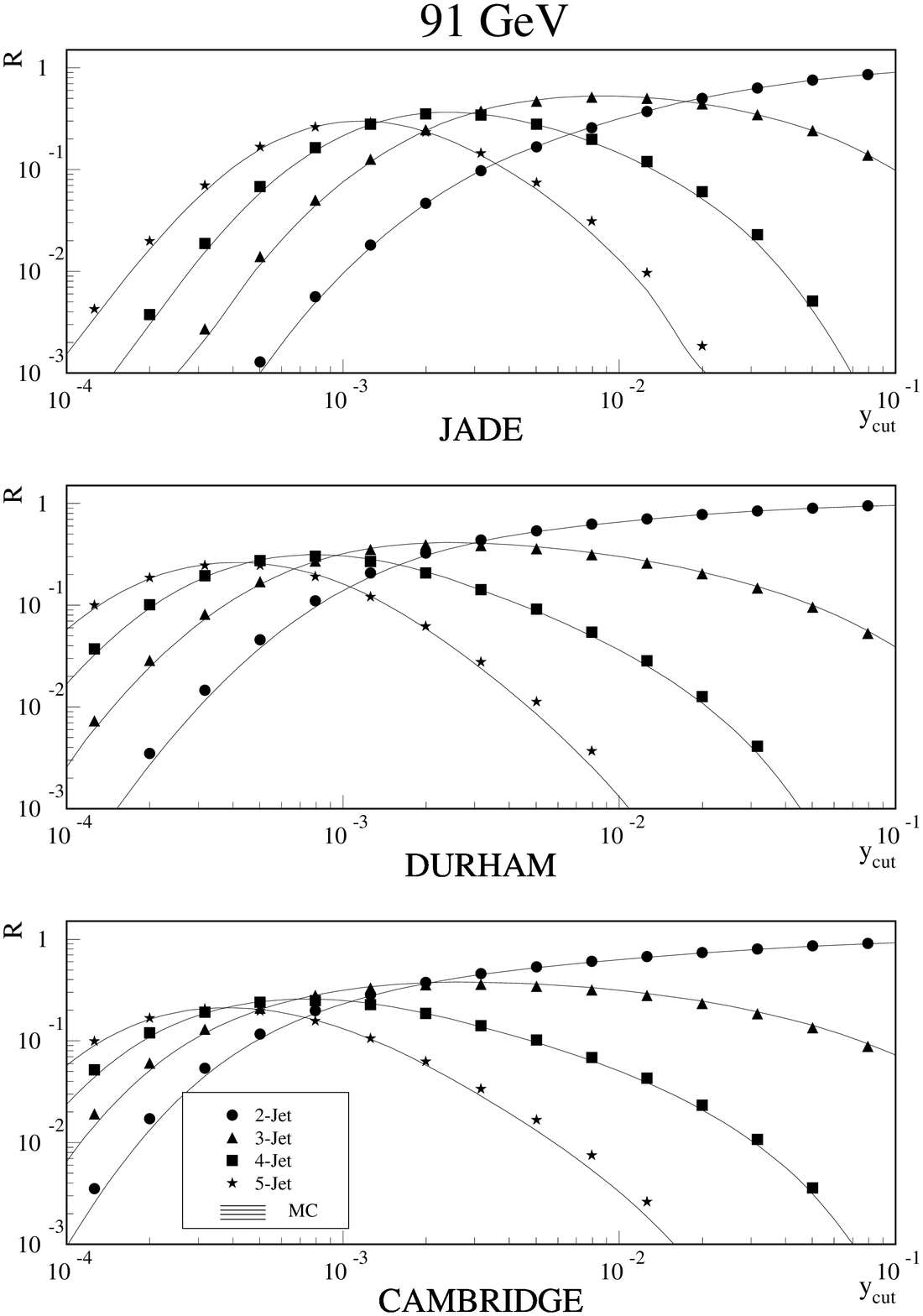,height=20cm} } 
  \end{center} 
  \vspace{0cm} 
  \myhangcaption{}{\label{fig:rates_91gev}Jet rates ($R$) at 91\gev\ as a 
  function of $y_{cut}$ compared to the prediction of  PYTHIA 6.1} 
\end{figure}

\begin{figure}[p] 
\vspace{18cm} 
\hspace{15cm} 
\begin{rotate}{90} 
    \mbox{ \epsfig{file=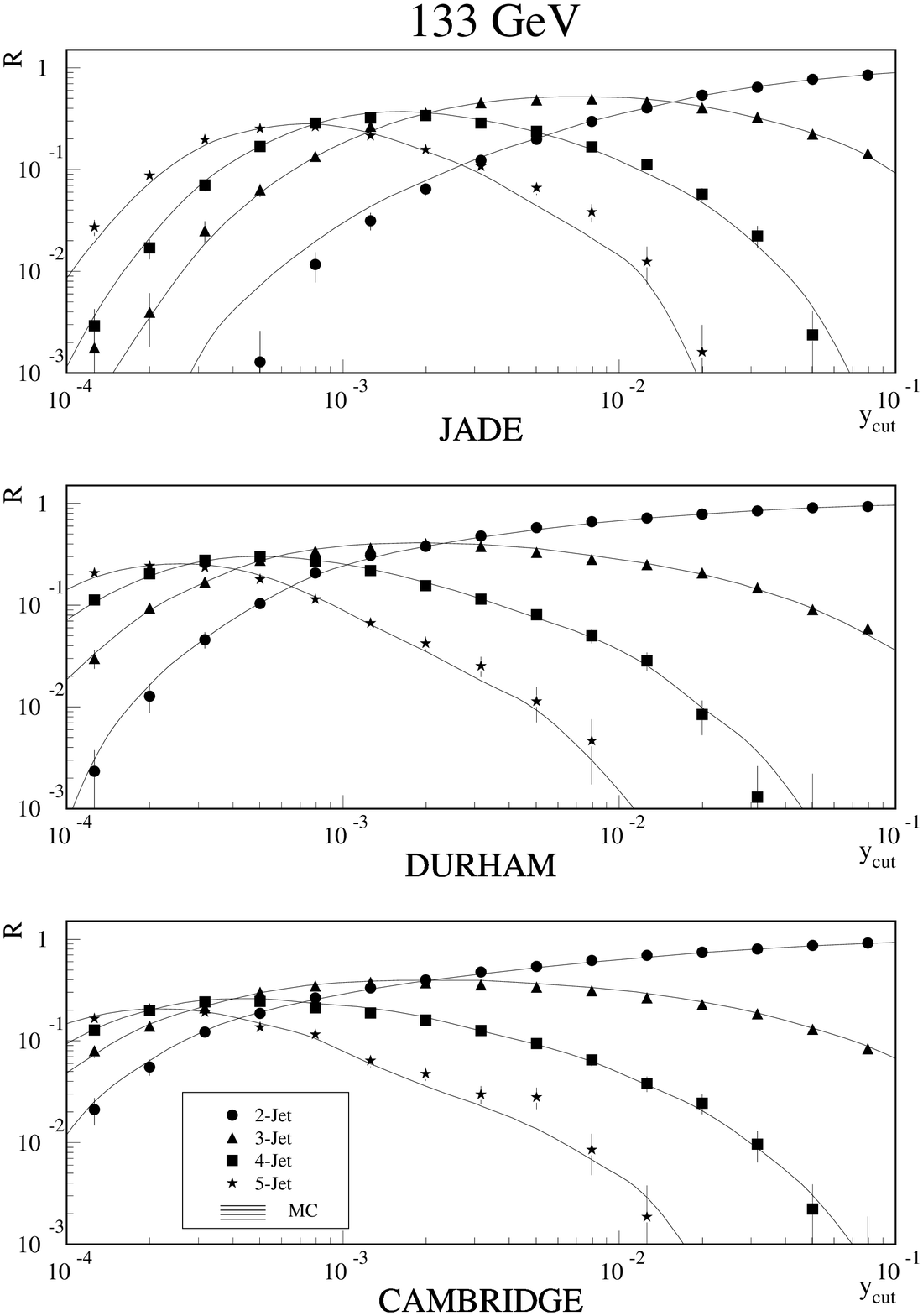,height=14cm} } 
    \mbox{ \epsfig{file=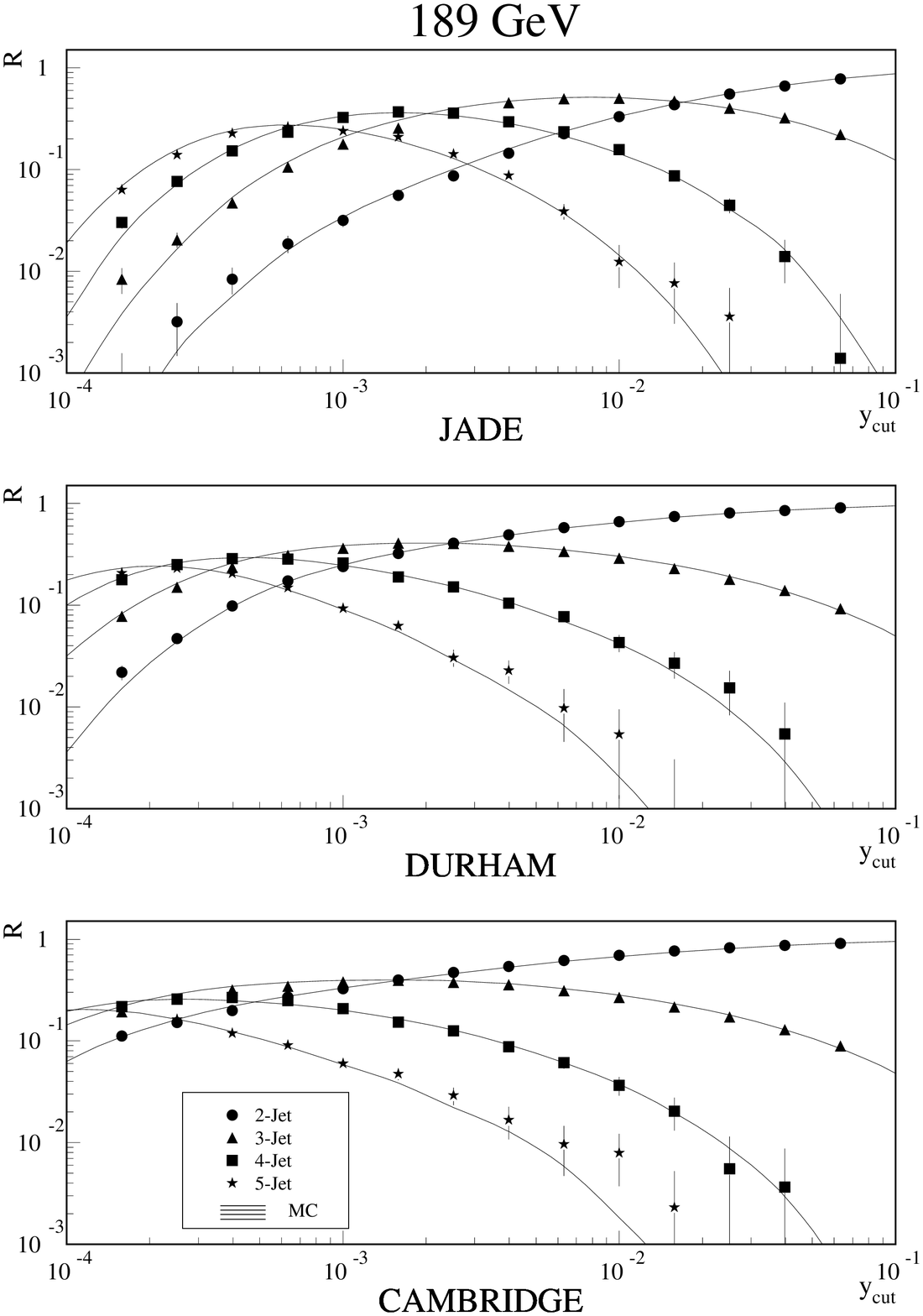,height=14cm} } 
\end{rotate} 
  \myhangcaption{}{\label{fig:rates_133-189gev}Jet rates ($R$) at 
  133 and 189\gev\ as 
  a function of $y_{cut}$ compared to the prediction of PYTHIA 6.1.} 
\end{figure}

\begin{figure}[p] 
\vspace{18cm} 
\hspace{15cm} 
\begin{rotate}{90} 
    \mbox{ \epsfig{file=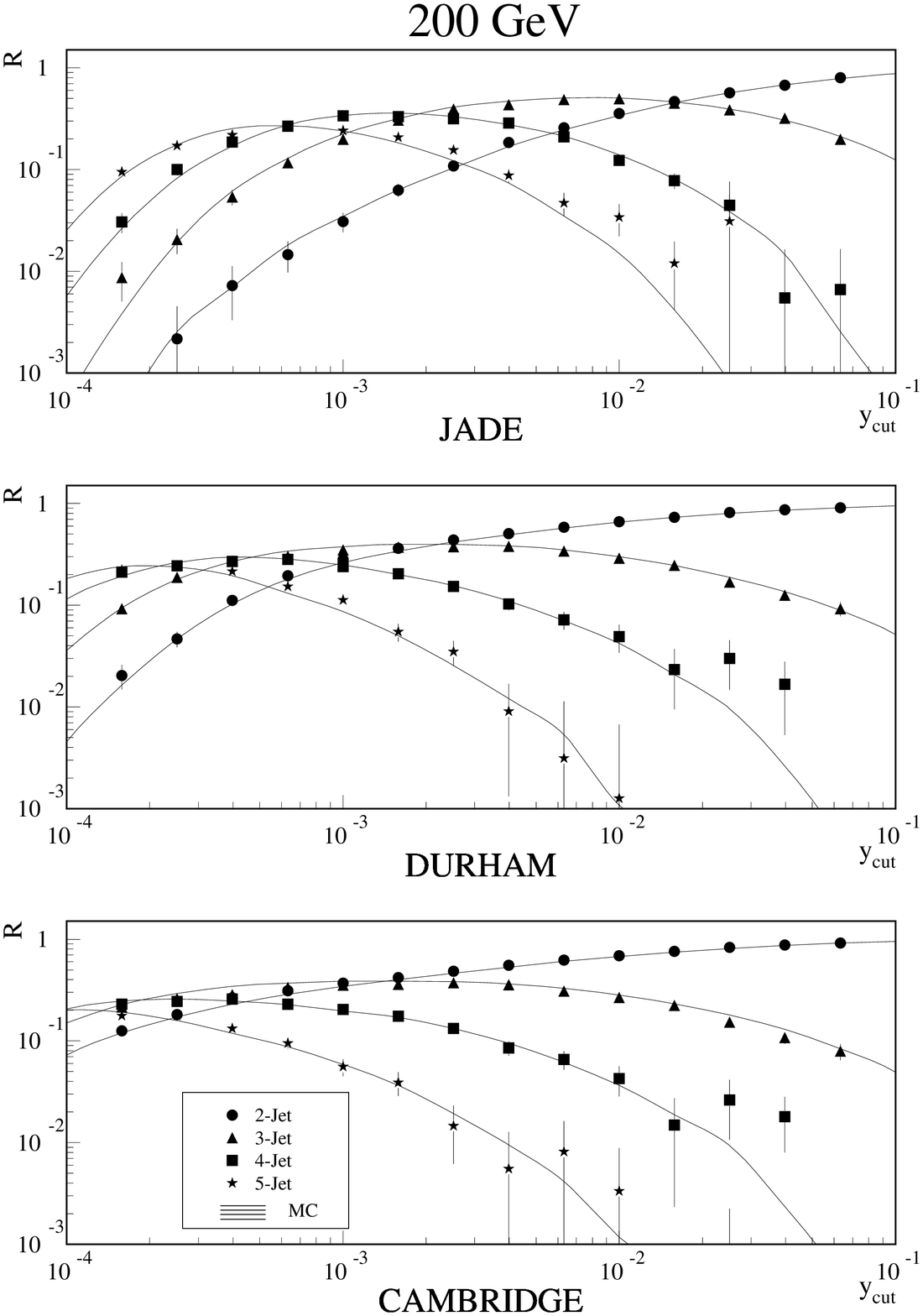,height=14cm} } 
    \mbox{ \epsfig{file=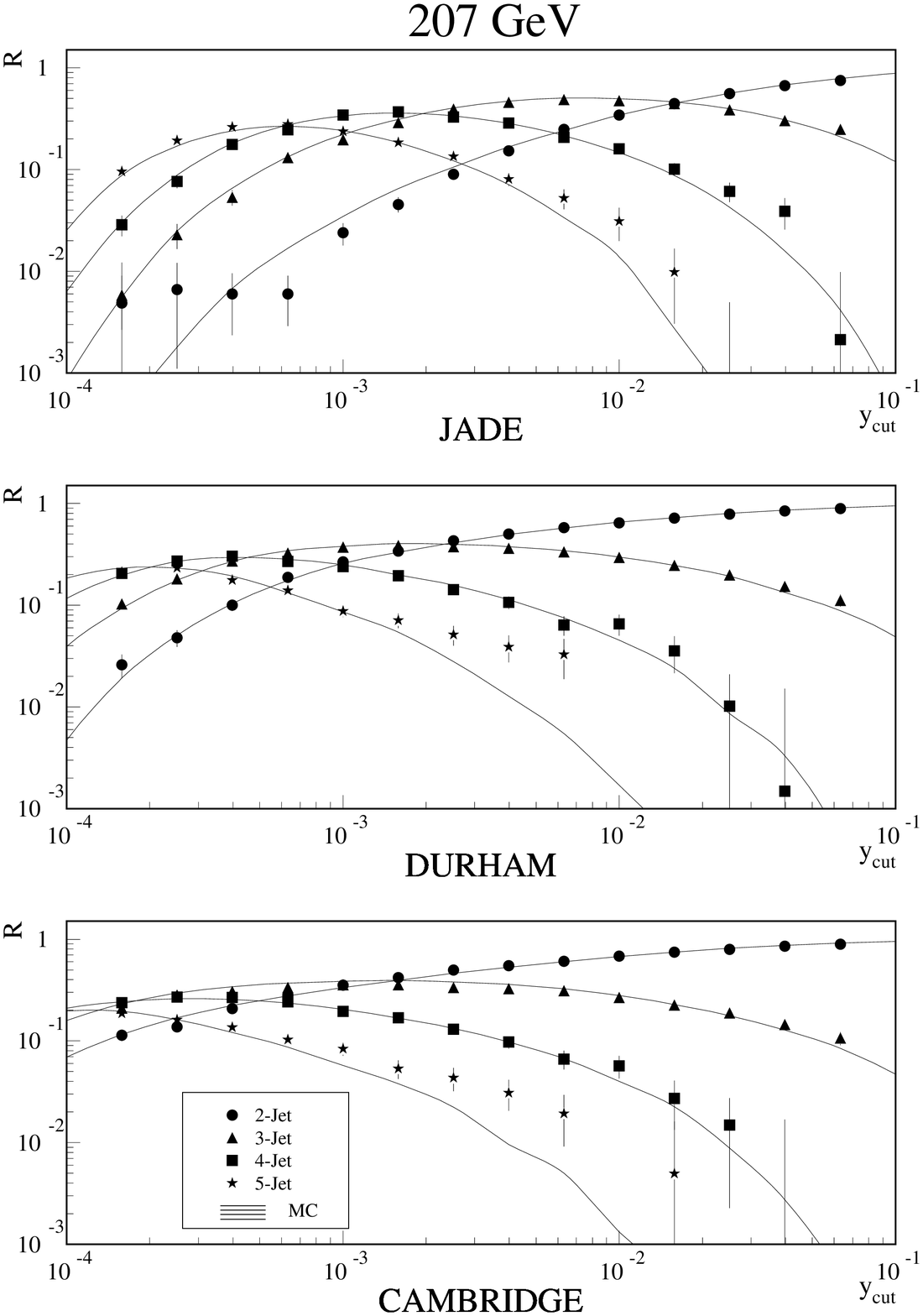,height=14cm} } 
\end{rotate} 
  \myhangcaption{}{\label{fig:rates_200-207gev}Jet rates ($R$) at 
  200 and 207\gev\ as 
  a function of $y_{cut}$ compared to the prediction of PYTHIA 6.1.} 
\end{figure}

\begin{figure}[p] 
  \begin{center} 
    \mbox{ \epsfig{file=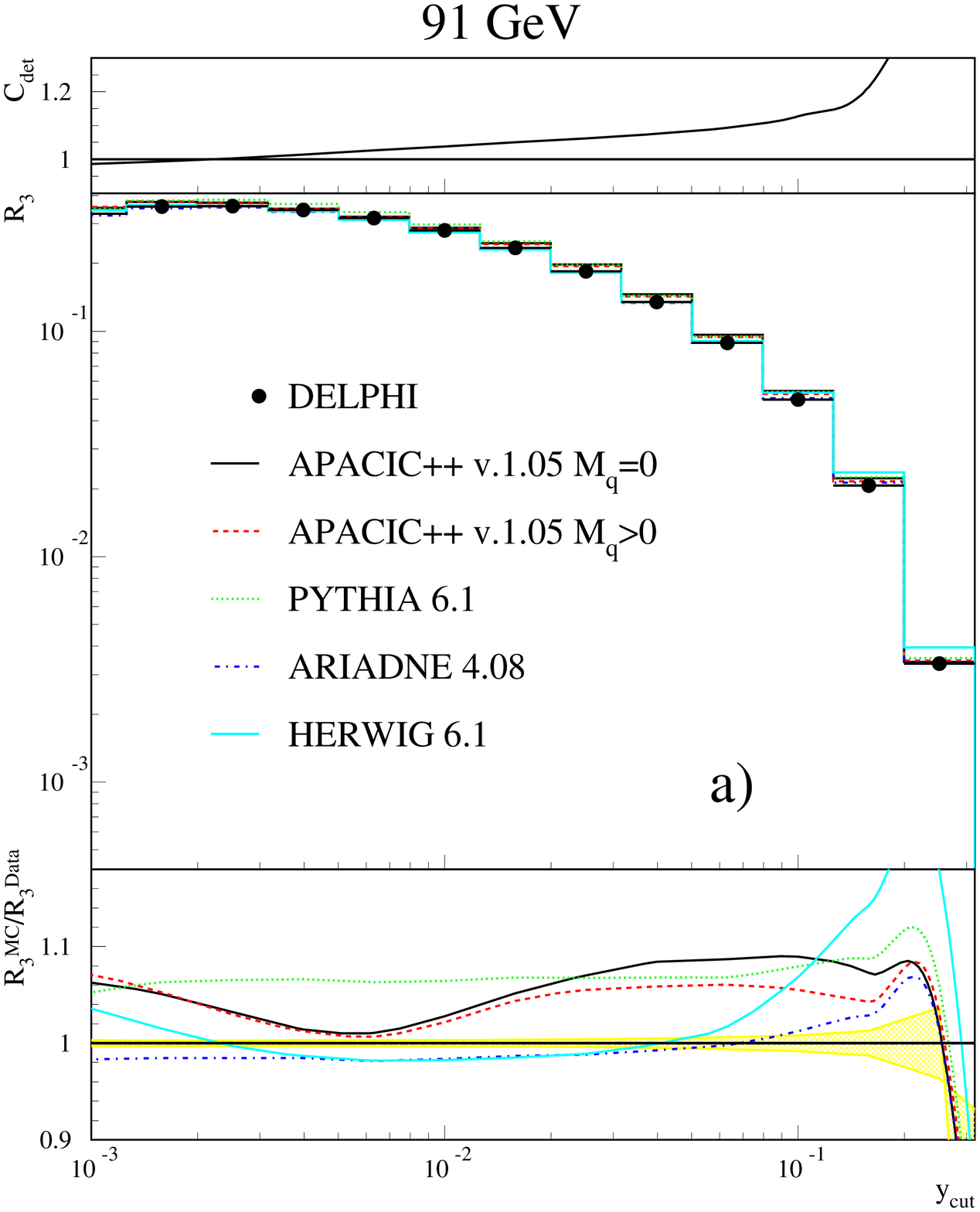,height=10.8cm} } 
    \mbox{ \epsfig{file=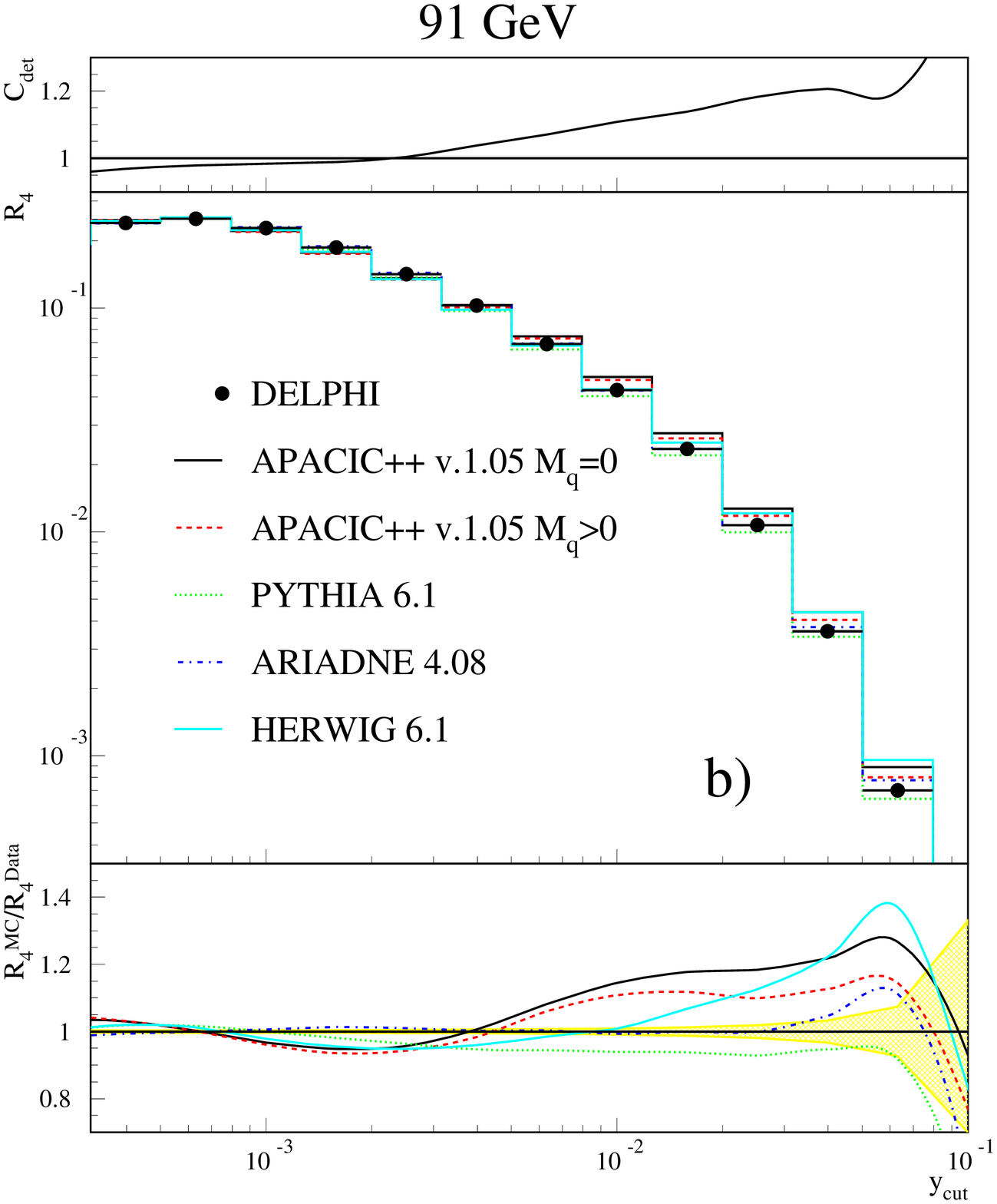,height=10.8cm} } 
  \end{center} 
  \vspace{0cm} 
  \myhangcaption{RCAM91}{\label{fig:r_cam}Jet rates determined with the 
  \cambridge\ algorithm. a) 3-jet rate at 91\gev. b) 4-jet rate 
  at 91\gev. The upper inset shows the corrections $C_{\mathrm{det}}$ applied
  to the data. The central plot shows the jet rates with 
  their statistical error in comparison with different Monte Carlo 
  predictions. The lower inset shows the jet rates normalised to the data. The
  band indicates the statistical and systematical uncertainty of the data.} 
\end{figure}

\begin{figure}[t] 
  \begin{center} 
     \mbox{ \epsfig{file=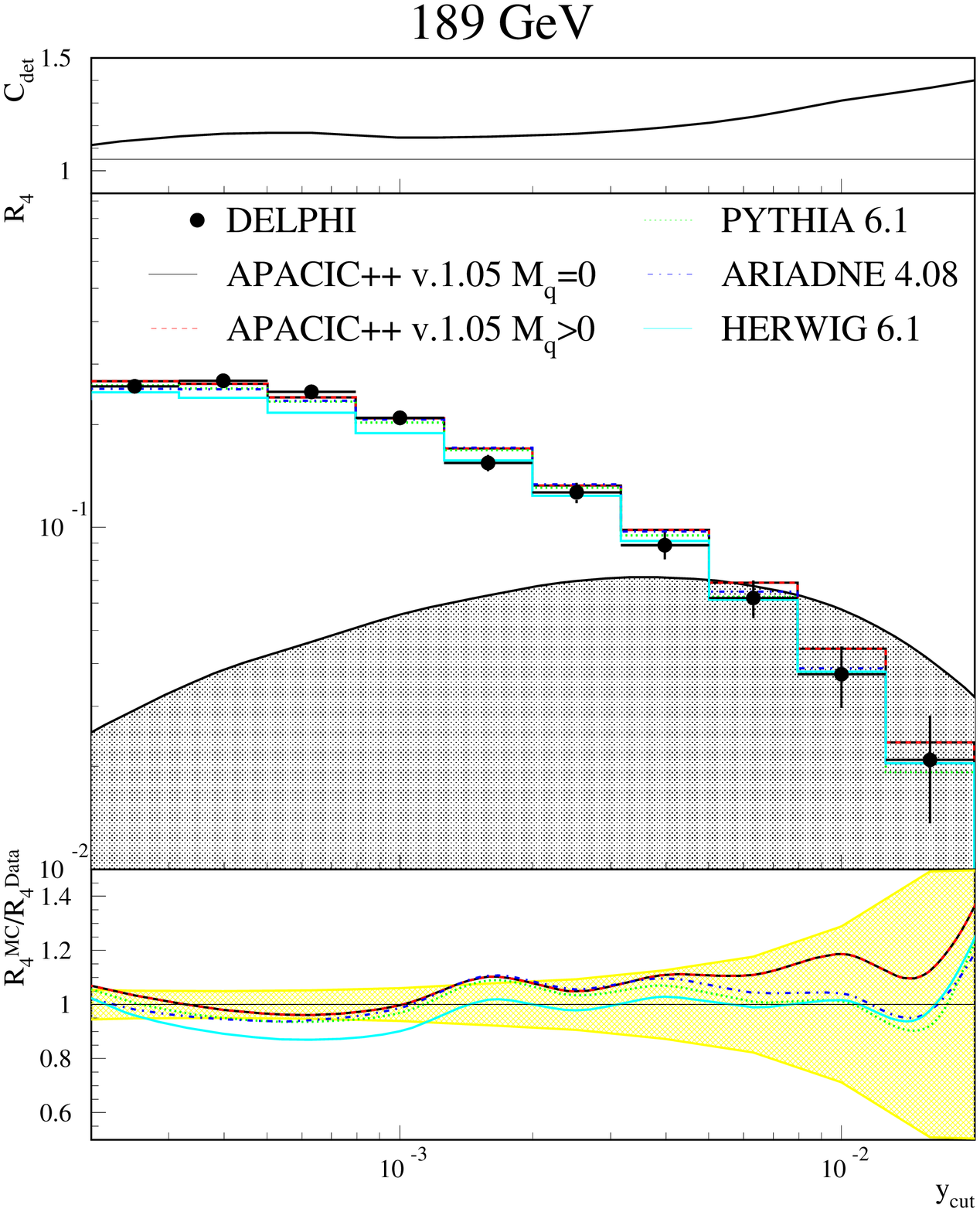,height=11.2cm} } 
  \end{center} 
  \vspace{0cm} 
  \myhangcaption{}{\label{fig:r_cam_189}
  4-jet rate at 189\gev\ determined with the \cambridge\ algorithm. The upper
  inset shows  the corrections $C_{\mathrm{det}}$ applied to the data.
  The central plot shows the jet rates with 
  their statistical error in comparison with different Monte Carlo 
  predictions. The grey area in the central plot shows the already 
  subtracted background of $WW$ and $ZZ$ events. The lower inset shows the 
  jet rates normalised to the data. The band indicates the statistical 
  and systematical uncertainty of the data.} 
\end{figure}

The remaining amount of four-fermion background is subtracted to obtain the
 final   data points given in 
\mbox{Figures \ref{fig:rates_91gev}, \ref{fig:rates_133-189gev} and 
  \ref{fig:rates_200-207gev}} 
showing the jet rates $R_2$, $R_3$, $R_4$ and $R_5$ as determined 
with the \jade, \durham\ and \cambridge\ jet algorithms at 91\gev\ 
and for a sample of \leptwo\ energies. 
Within errors, the 2-, 3-, and 4-jet rates show a good overall agreement
at all energies  with the generator predictions tuned to data at the $Z$
 resonance. Figures \ref{fig:r_cam} and \ref{fig:r_cam_189} show a 
detailed comparison between 
\cambridge\ jet rates and Monte Carlo predictions. 
Several models, tuned to \delphi\ data, are available 
\cite{Abreu:1996na,Ballestrero:2000ur} and are used within this analysis: 
\begin{itemize} 
 \item 
    \pythia~6.1 is a parton-shower model with explicit angular ordering,
    followed by string fragmentation~\cite{Sjostrand:1995iq}. 
  \item 
    \ariadne~4.08 performs a colour-dipole cascade,  followed by 
    string fragmentation~\cite{Lonnblad:1991rx}. 
  \item 
    \herwig~6.1 is a coherent parton-shower model, followed by 
    cluster fragmentation~\cite{Marchesini:1996vc}. 
  \item 
    \apacic\ performs a massive leading-order (LO) matrix element (ME) 
    calculation for 
    3-, 4-, and 5-jet final states, 
    matched to a parton-shower and followed by Lund 
    string fragmentation~\cite{Kuhn:2000dk,Krauss:1999fc,Krauss:1999mt}. 
    The \apacic\ parameters have been tuned to DELPHI data 
    measured at the $Z$ resonance.
\end{itemize} 

The precise \lepone\ data in particular allow a critical judgment of the 
precision of tuned Monte Carlo models \cite{Ballestrero:2000ur}. 
The parton-shower model \pythia\ tends to 
overestimate the 3-jet rate and to underestimate the 4-jet rate at 
large \ycut, see also Figure \ref{fig:rates_91gev}. To cure the 
lack of multijet events a calculation of the underlying matrix 
elements has been performed. \apacic\ 
also tends to overestimate the 3-jet rate but predicts more 4-jet 
events at small \ycut. By taking quark mass effects into account 
the agreement with the data improves somewhat. The parton-shower generator 
\herwig\ also gives an acceptable 
agreement with the data. The best overall agreement is obtained with
the colour-dipole model \ariadne. At \leptwo\ energies the
deviations are  
obscured by the larger statistical errors. Within errors all 
models show good agreement with the data. Note that the errors 
shown are statistical only and that neighbouring bins are 
correlated. Considering the experimental errors and model 
uncertainties, no significant excess of multijet events at higher 
energies is observed.

 
%
%
 
%
 
\clearpage 
 
\section{Determination of \boldmath\as} 
\label{alphas} 
The strong coupling constant, \as, is determined from the four-jet rate, 
by fitting an \as-dependent QCD 
prediction folded with a hadronisation correction to the data. 
 
\subsection{NLO predictions}
The next--to--leading--order (NLO) expression of the four-jet rate is given
by:  
\begin{equation} 
  \label{eq:r4_nlo} 
       R_4(y) = B(y) \cdot \alpha_s^2 + 
        [C(y) +2B(y)\cdot b_0\cdot \ln{x_{\mu}}] \cdot \alpha_s^3 + \ldots \, . 
\end{equation} 
Where $x_\mu=\mu^2/Q^2$, $\mu$ being the renormalisation scale, 
$Q$ the centre-of-mass energy of the event, $b_0=(33-2n_f)/12\pi$ 
and $n_f$ the number of active flavours. \eqn{eq:r4_nlo} shows the 
explicit 
dependence of $R_4$ on 
the renormalisation scale $\mu$. The coefficients $B(y)$ and 
$C(y)$ for the \durham\ and the \cambridge\ algorithms are obtained 
by integrating the massless matrix elements for $e^+e^-$ 
annihilations into four-parton final states, performed by the NLO 
generator \debrecen\ \cite{Nagy:1998bb,Nagy:1997yn}.
The $R_4$ results obtained with the \jade\ algorithm are not used
for the $\alpha_s$ determination because of phantom jets and larger
hadronisation corrections. 
 
%
 
\begin{figure}[tbh] 
\unitlength1cm 
 \unitlength1cm 
 \begin{center} 
 %
 \begin{minipage}[t]{7.5cm} 
           \mbox{\epsfig{file=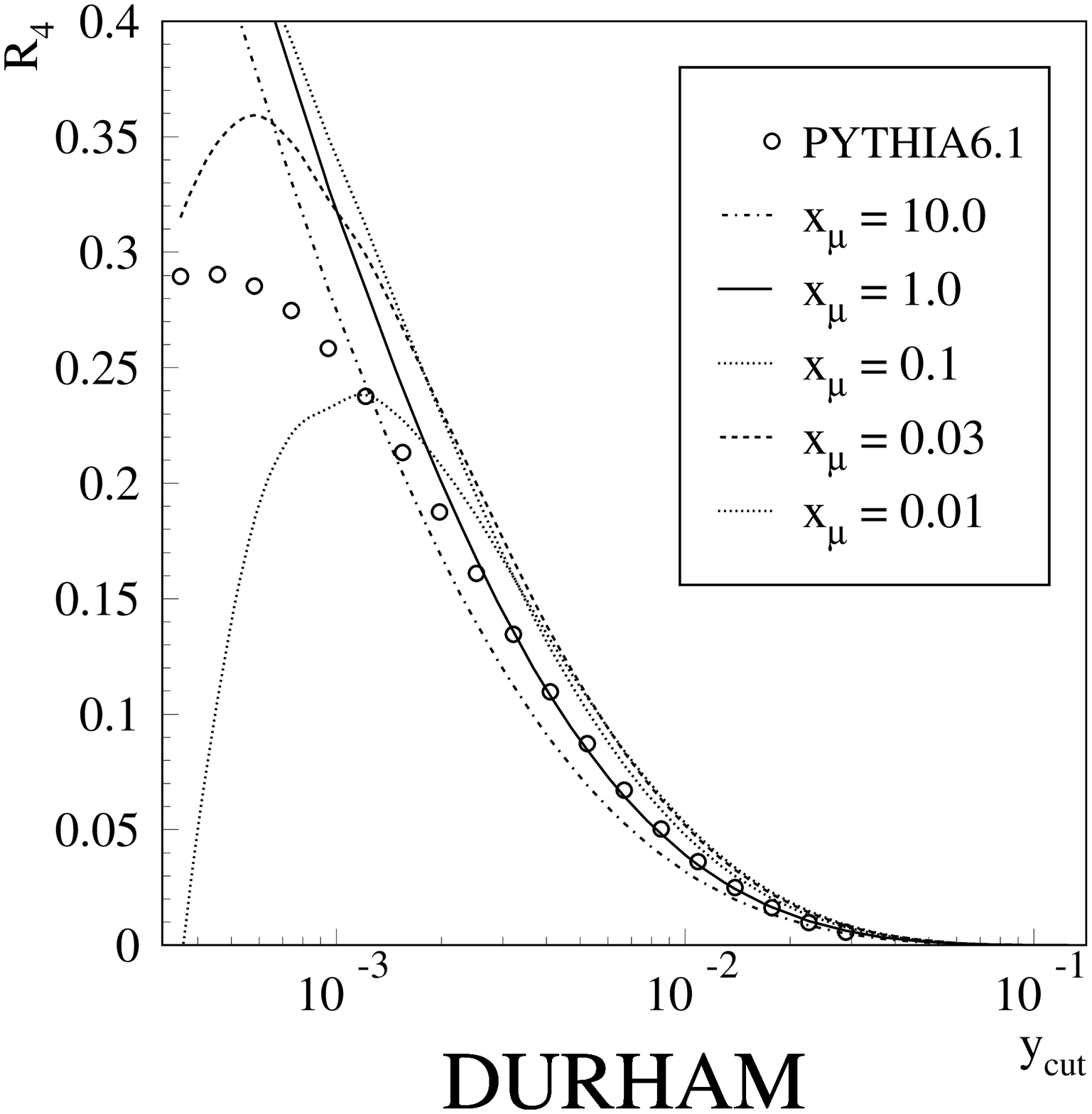,width=8.0cm}} 
 \end{minipage} 
 \hspace{.3cm} 
 \begin{minipage}[t]{7.5cm} 
           \mbox{\epsfig{file=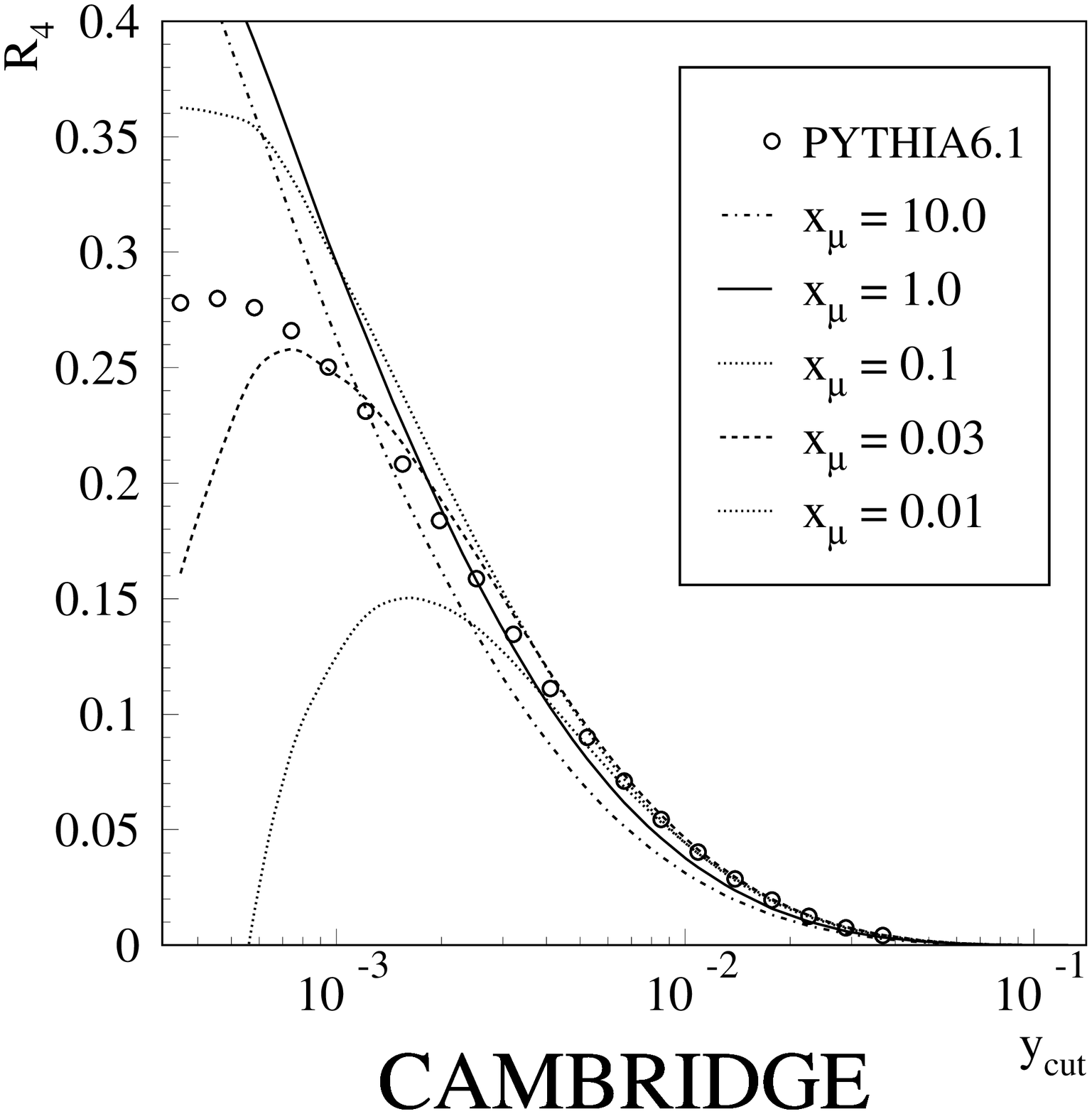,width=8.0cm}} 
 \end{minipage} 
\end{center} 
\myhangcaption{LO_NLO}{\label{fig:some_scales_as}Predictions (left for DURHAM,
   right for CAMBRIDGE) of 
   the four-jet rate $R_4$ at 91 GeV using DEBRECEN and \eqn{eq:r4_nlo} 
   for various values of $x_\mu$ at fixed $\alpha_s=0.118$. 
   Illustrated is the change of the prediction with varying $x_\mu$. The
   analytical  
   calculations are compared to the parton level prediction using the PYTHIA
   generator. }  
\end{figure}

\fig{fig:some_scales_as} shows the dependence of $R_4$ on 
\xmu. For small values of \xmu\ the overshoot of the NLO 
expression changes into an underestimate of the observed/measured $R_4$. 
Thus small values of \xmu\ are expected 
when fitting the data, suggesting important contributions of 
higher-order corrections of \oasss. 
 

%

\subsection{Hadronisation} 
Before comparing \eqn{eq:r4_nlo} with the data and fitting its 
parameters \as\ (and \xmu), the transition of coloured partons 
into colourless hadrons has to be accounted for. This transition 
has been simulated using Monte Carlo fragmentation models. For 
each centre-of-mass energy the QCD prediction is 
multiplied by the hadronisation correction \begin{equation} 
C_{\mathrm{had}}(E_{\mathrm {cm}})= 
        \frac{f_{\mathrm {had}}^{\mathrm {Sim.}}(E_{\mathrm {cm}})} 
             {f_{\mathrm {part}}^{\mathrm {Sim.}}(E_{\mathrm {cm}})} 
   \quad\mbox{,} 
\label{eq_hadc} \end{equation} where $f_{\mathrm {had}}^{\mathrm 
{Sim.}}(E_{\mathrm {cm}})$ ($f_{\mathrm {part}}^{\mathrm 
{Sim.}}(E_{\mathrm {cm}})$) is the model prediction on the hadron 
(parton) level at the centre-of-mass energy $E_{\mathrm {cm}}$. 
The parton level is defined as the final state of the parton 
shower created by the Monte Carlo event generation.

\begin{figure}[t]
\unitlength1cm 
 \unitlength1cm 
 \begin{center} 
 \begin{minipage}[t]{7.0cm} 
           \mbox{\epsfig{file=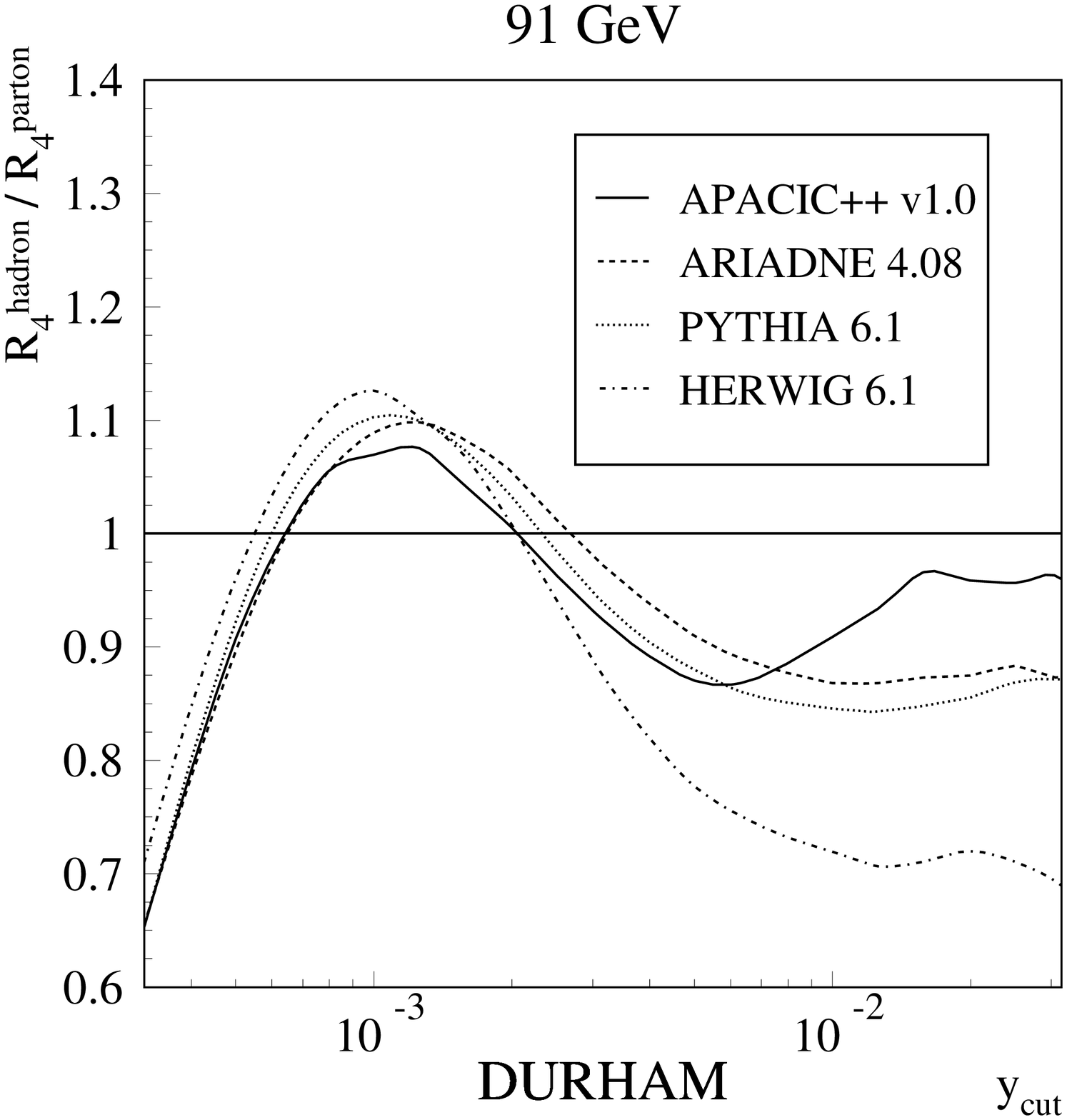,width=7.5cm}} 
 \end{minipage} 
 \hspace{.3cm} 
 \begin{minipage}[t]{7.0cm} 
           \mbox{\epsfig{file=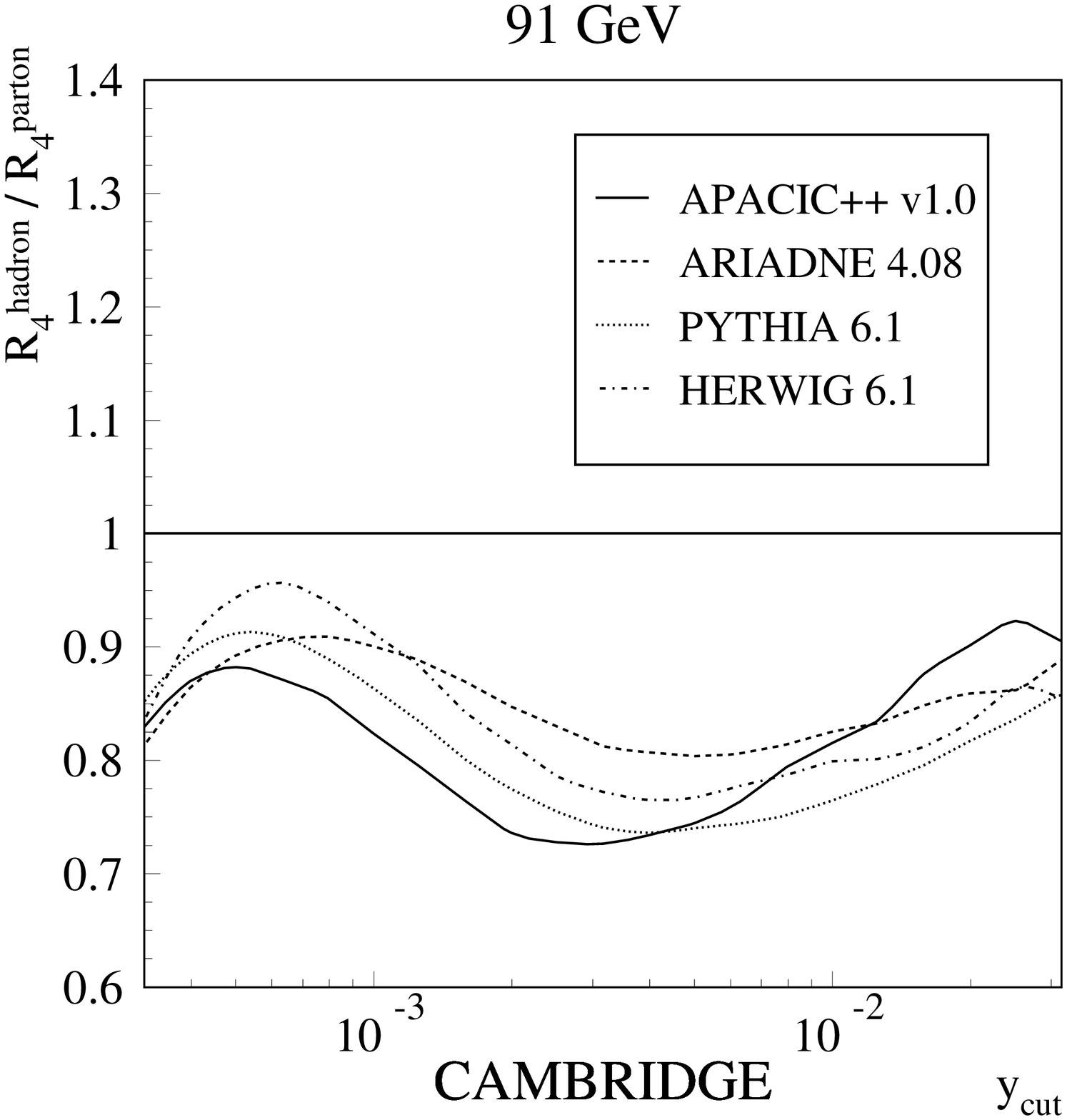,width=7.5cm}} 
 \end{minipage} 
\end{center} 
\myhangcaption{HADCOR}{\label{fig:hadcor_allmc}Distribution of the 
  hadronisation corrections to the four-jet rate. The plots show 
  the ratio of the four-jet rates after and before simulation of the 
  fragmentation, evaluated with different Monte Carlo models.} 
\end{figure}

The matching of ME calculations with a parton shower within 
\apacic\ allows the tuning, performed at \lepone\ 
energies, to be extrapolated. Thus \apacic\ is the only ME 
generator available at \leptwo\ energies. 
Therefore \apacic\ is taken as the reference model. The scatter of 
results in \as, when using different Monte Carlo generators is 
added to the systematic error. \fig{fig:hadcor_allmc} shows the 
ratios between hadron level and parton level as a function of 
\ycut\ from different generators. Using the \cambridge\ algorithm 
the ratio $R_4^{hadron}/R_4^{parton}$ shows a weaker \ycut\ 
dependence than the same ratio determined by using \durham. 
 
\subsection{\label{subsec:xmu}Dependence on the 
renormalisation scale $\boldmath\mu$} 
%
The explicit dependence of \as\ derived from \eqn{eq:r4_nlo} on 
the renormalisation scale \xmu\ arises from the truncation of the 
perturbative series after a fixed number of orders. Within 
perturbative QCD \xmu\ is an arbitrary parameter. A conventional 
scale setting called ``physical scale'' is the choice $x_\mu=1$. 
However, several other proposals for evaluating the 
renormalisation scale are available in the literature. Two of them 
are investigated within this analysis: 
\begin{itemize} 
\item Method of effective charges (ECH) \cite{Grunberg:1984fw}:\\ 
  In \oass\ perturbation theory, the ECH scale value has to be chosen in such 
  a way that the third-order term vanishes: 
  \begin{equation} 
    \label{eq:def:ech} 
    C(y) +2B(y)\cdot b_0\cdot \ln{x_{\mu}} = 0 \, . 
  \end{equation} 
\item Principle of minimal sensitivity (PMS) \cite{Stevenson:1981vj}:\\ 
  The PMS optimisation amounts to the determination of the renormalisation 
  scale value, which minimises the sensitivity of the theoretical prediction 
  with respect to its variation: 
  \begin{equation} 
    \label{eq:def:pms} 
    \frac{d}{dx_\mu}\left[C(y) +2B(y)\cdot b_0\cdot 
                                   \ln{x_{\mu}} \right] = 0 \, . 
  \end{equation} 
\end{itemize} 
Within both theoretical scale-setting methods the scale \xmu\ is a 
function of \ycut. The uncertainty of the scale is conventionally 
estimated by a scale variation within an ad hoc chosen range. 
 
\begin{figure}[b] 
\unitlength1cm 
 \unitlength1cm 
 \begin{center} 
 \begin{minipage}[t]{10.0cm} 
  \mbox{ \epsfig{file=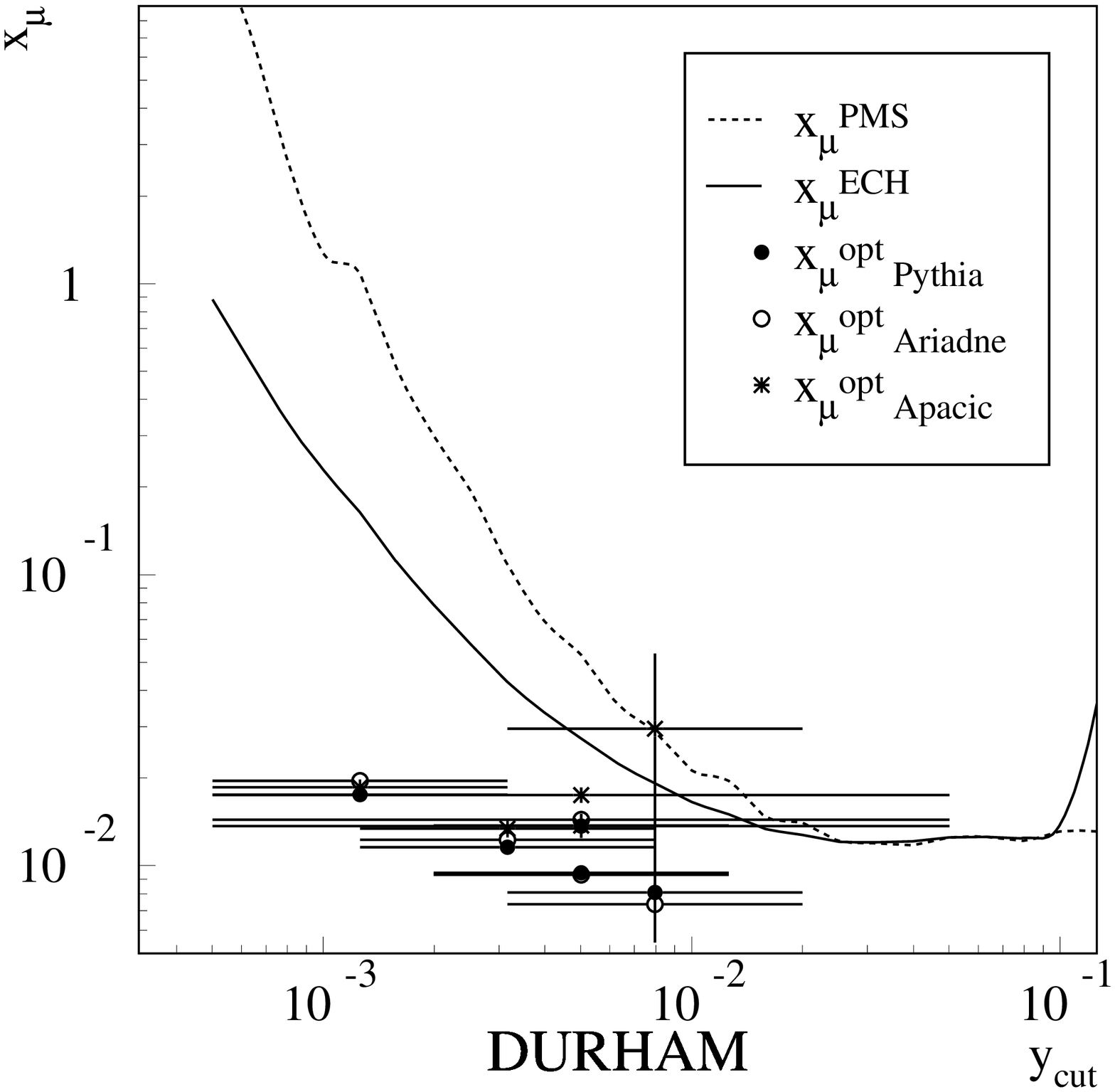,width=10.cm} } 
 \end{minipage} 
 \hspace{.3cm} 
 \begin{minipage}[t]{10.0cm} 
  \mbox{ \epsfig{file=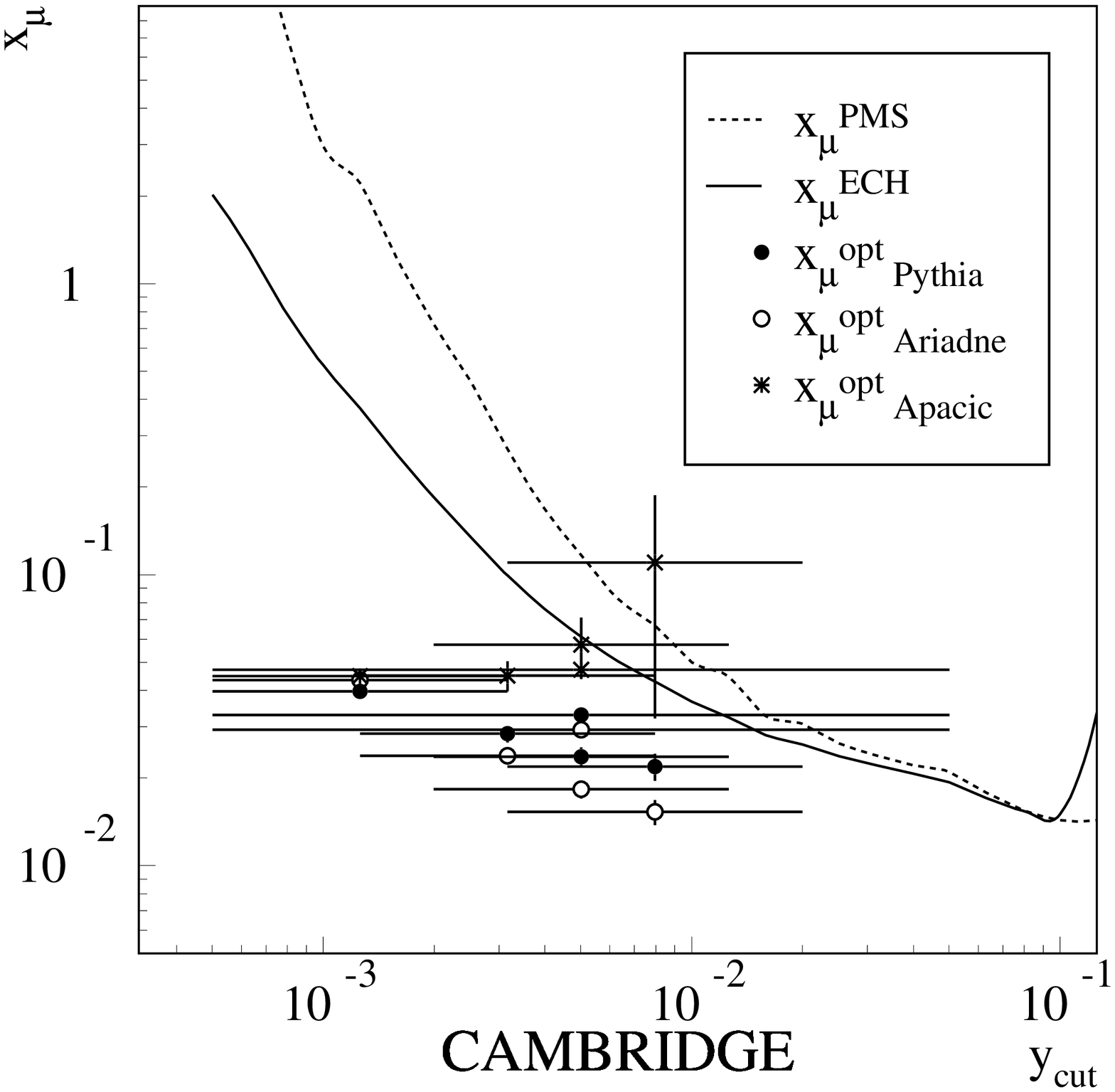,width=10.cm} } 
 \end{minipage} 
\end{center} 
  \myhangcaption{Scales} {\label{fig:opt-scales}Optimised renormalisation 
  scales: The lines show the $y_{cut}$  dependence for theoretically optimised
  scales at $E_{\mathrm {cm}}=91$ GeV. The dots give results for
  experimentally optimised scales. The error bars in the horizontal direction
  indicate the fit range, the different symbols represent hadronisation  
  corrections applied by different Monte Carlo models.} 
\end{figure}

In perturbative QCD the \xmu\ dependence of the prediction for an 
observable $R$ would vanish in the all orders limit only. It has 
been shown in \cite{Abreu:2000ck} that an excellent description of 
precise $m_Z$ data can be obtained by fitting  simultaneously \as\ and \xmu.  
In the same way a simultaneous fit of \as\ and \xmu\ to the jet rates was
performed to account for the missing higher-order calculations. 
The fitted scale is called the experimentally optimised scale 
\xmuopt. The results of the scale-setting methods are shown in 
\fig{fig:opt-scales}. Experimentally optimised scales for different 
fit ranges (indicated by the error bars in $y_{\mathrm {cut}}$ 
-direction) and for several hadronisation models are compared with 
the theoretical scale evaluations. The fit ranges for \xmuopt\ are 
varied between $y_{\mathrm {cut}} = 0.05$ and $y_{\mathrm {cut}} = 
0.0005$. Below $y_{\mathrm{cut}} \simeq 0.0005$ the perturbative 
expansion is expected to become invalid, above 
$y_{\mathrm{cut}}\simeq 0.05$ the number of events entering $R_4$ 
becomes too small to perform the fit. 
For $y_{cut} > 0.001$ small values of \xmu\ are preferred and for 
 $y_{cut}$ near $0.01$ the theoretical scale evaluations are of the same
magnitude as the experimentally optimised scales.

\clearpage 
 
\subsection{Fits to \lepone\ data and a precise measurement of \boldmath\asmz} 
%
As discussed above, for each measurement of \as\ the 
renormalisation scale has to be chosen. To determine the 
experimentally optimised scale a two-parameter fit of 
\eqn{eq:r4_nlo}\ with \as\ and \xmuopt\ as free parameters is 
performed to the four-jet rate. 
\tab{tab:opt-scales} shows the results for \xmuopt. 
 
\begin{table}[htbp] 
  \begin{center} 
    \begin{tabular}{|c||c|c|} 
      \hline 
      algorithm  & fit range               & \xmuopt \\ \hline \hline 
      \durham    & $0.001 - 0.01$ & 0.015 \\ \hline 
      \cambridge & $0.001 - 0.01$ & 0.042 \\ \hline 
    \end{tabular} 
    \myhangcaption{xxx}{\label{tab:opt-scales}Experimentally optimised
    scales.}  
  \end{center} 
\end{table}

\begin{figure}[p] 
\unitlength1cm 
 \unitlength1cm 
 \begin{center} 
 \begin{minipage}[t]{7.5cm} 
  \mbox{ \epsfig{file=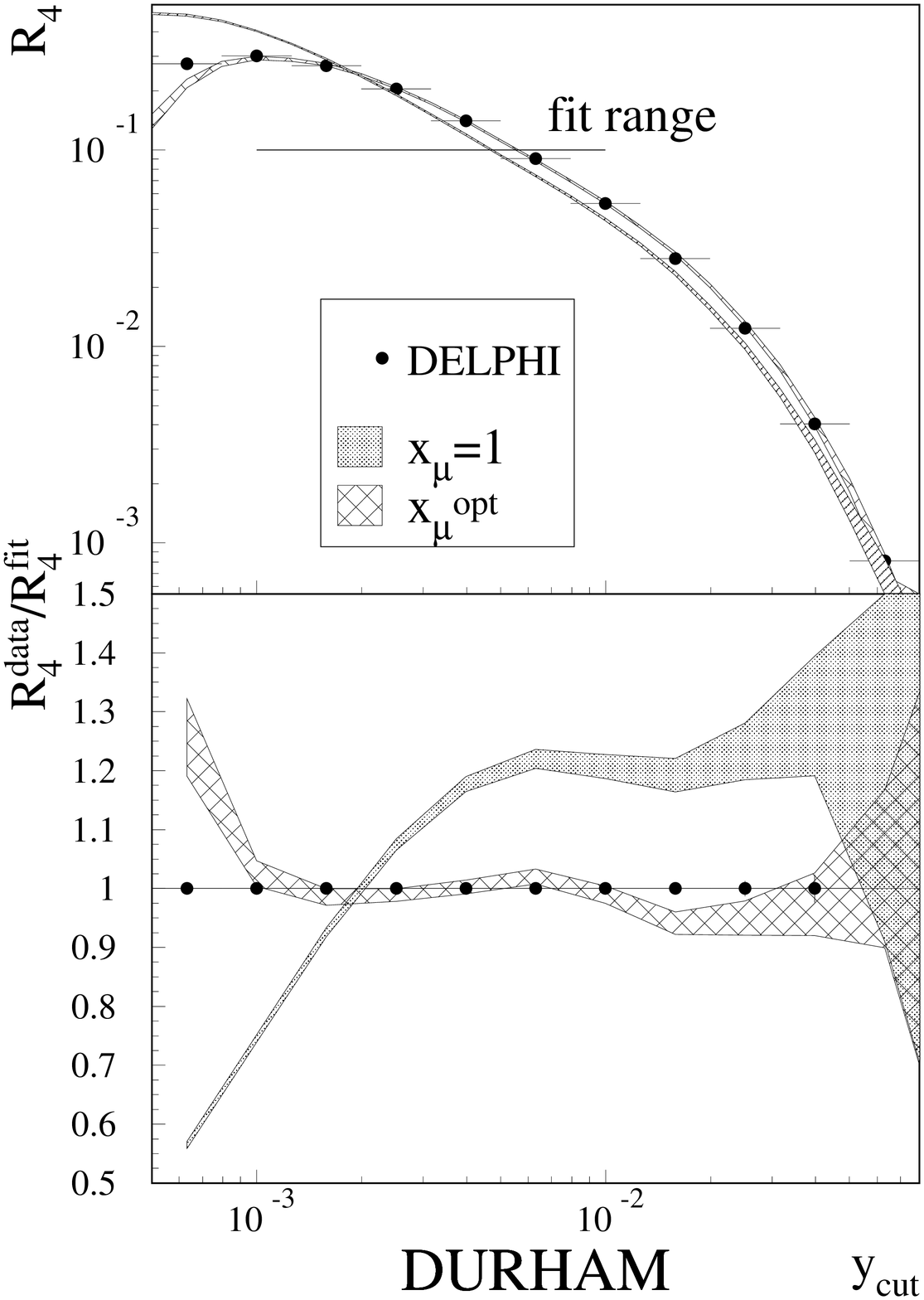,width=8.0cm} } 
 \end{minipage} 
 \hspace{.3cm} 
 \begin{minipage}[t]{7.5cm} 
  \mbox{ \epsfig{file=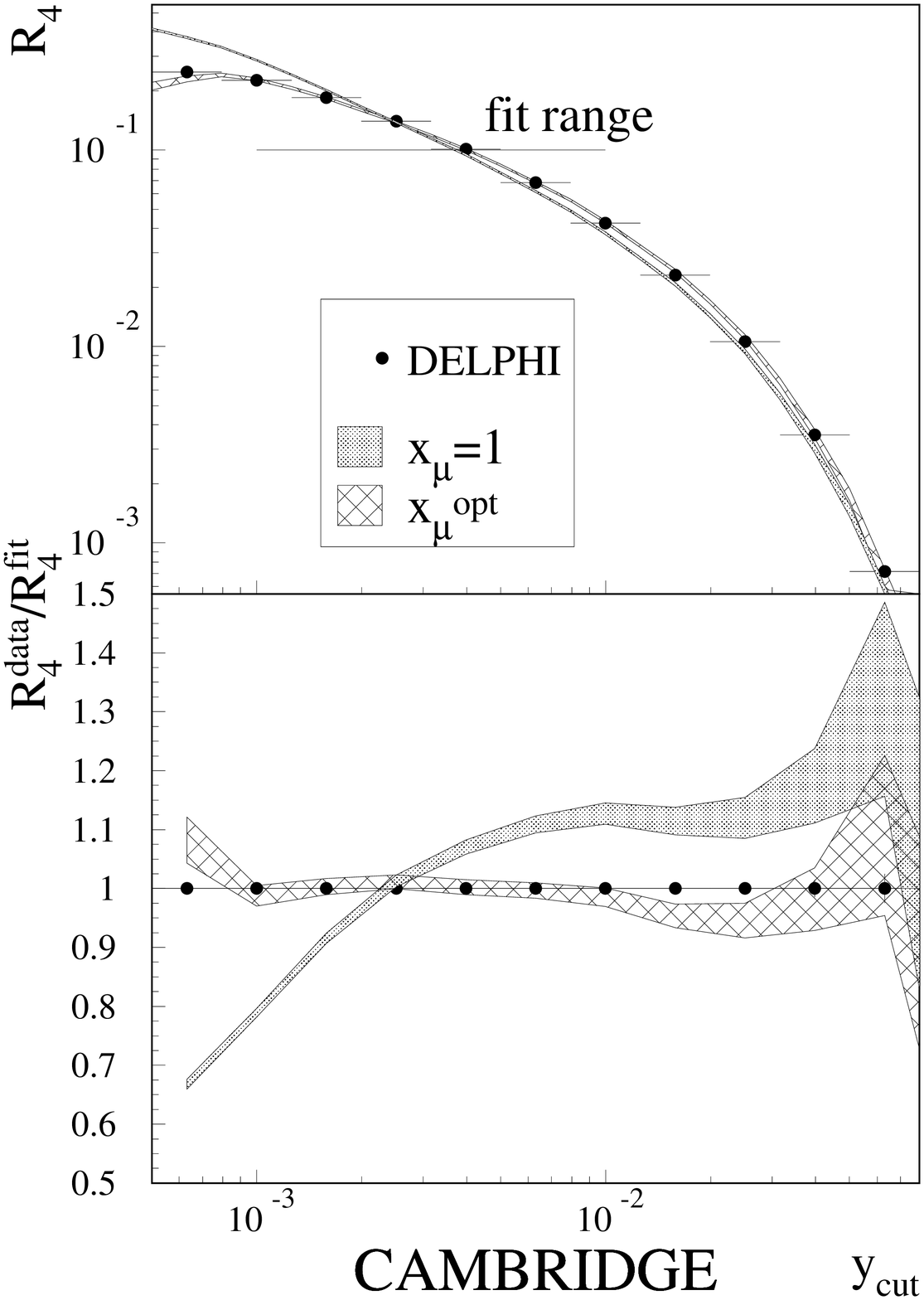,width=8.0cm} } 
 \end{minipage} 
\end{center} 
  \myhangcaption{Fits an \vjr} {\label{fig:fit_durcam}Fits to the 
    four-jet rate $R_4$ measured at the Z resonance using different scale
    evaluation methods. Top: the distributions. The hatched curve shows
    the results for the experimentally optimised scales.
    Bottom: the ratio
    $R_4^{\mathrm{data}}/R_4^{\mathrm{fit}}$.  The grey bands  
    show fit results with the physical scale ($x_\mu=1$), the cross-hatched
    bands for  
    experimentally optimised scales (\xmuopt).} 
\end{figure}

\fig{fig:fit_durcam} shows results of fits using Eq.~\ref{eq:r4_nlo} and the
fit ranges given in Table~\ref{tab:opt-scales} for DURHAM and CAMBRIDGE and
for both physical ($x_{\mu}=1$) and experimentally optimised scales.
While the fit with experimentally optimised 
scales results in a good agreement with the data over two orders 
of magnitude in \ycut, the fit results with physical scale show a 
\ycut\ dependence inconsistent with the measurement. 
 
\begin{figure}[bthp] 
  \begin{center} 
  \mbox{ \epsfig{file=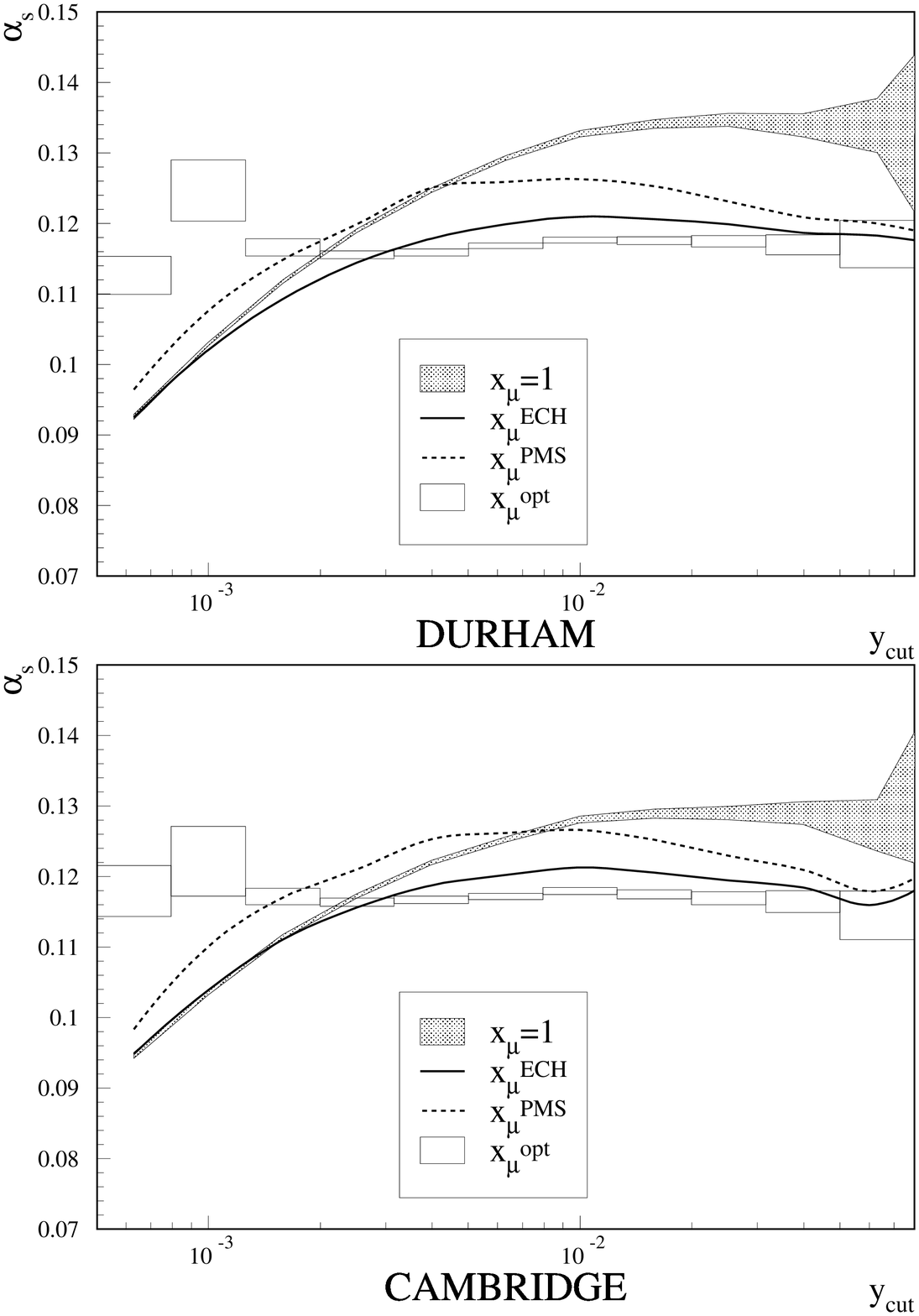,width=10.0cm} } 
  \end{center} 
  \myhangcaption{Fits an \vjr} {\label{fig:as_ycut}Dependence of \as\ 
   on \ycut: The grey bands give fit results with physical scale ($x_\mu=1$), 
   the lines with theoretically optimised scales (ECH, PMS) and the rectangles 
   with experimentally optimised scales (\xmuopt).} 
\end{figure}

\fig{fig:as_ycut} presents results of the $\alpha_s$ fits as a function of 
$y_{\mathrm cut}$ for different scale evaluation
methods.
For fits with physical scale the
resulting $\alpha_s$ values show a strong dependence on the choice of
$y_{\mathrm cut}$. Within the investigated range \as\ varies 
from about 0.1 to 0.13.
Theoretically optimised scales (ECH 
and PMS) improve the situation, but for small values of \ycut\, 
where the theoretical scales increase, \as\ shows again a strong 
dependence on \ycut. 
Choosing experimentally optimised scales 
cures the problem. With this choice the \as\ results are 
independent of \ycut, and furthermore results for \durham\ and 
\cambridge\ are in good agreement.
Experimentally optimised scales are therefore considered as an 
accurate tool to perform a consistent measurement of the strong 
coupling from four-jet rates.  

The jet rate data, as shown, for instance, in~\fig{fig:rates_91gev}, 
are highly correlated. Therefore a second fit is performed to just 
one single bin in \ycut\ with \as\ as the only free parameter 
using the fixed scales of Table \ref{tab:opt-scales}. This final 
fit is performed at $y_{\mathrm {cut}} = 0.0063$ for both the 
 \durham\ and \cambridge\ algorithms. As shown in \fig{fig:as_ycut} the fit
results are very stable in the vicinity of this \ycut\ value.

The total error on \asmz\ is estimated by considering the 
following experimental and theoretical uncertainties: 
\begin{itemize} 
\item Variations of the track and event cuts given in \tab{cuts}: $N_{\mathrm 
    {charged}} \geq (7\pm 1)$, $E_{\mathrm {tot}} \geq (0.50\pm 0.05)\cdot 
  E_{\mathrm {cm}}$ and $(25^\circ\pm 5^\circ) \leq \theta_{\mathrm {thrust}} 
  \leq (155^\circ\pm 5^\circ)$. 
\item In order to account for a remaining dependence on \ycut, the working 
  point is varied in the range $0.0016 \leq y_{\mathrm {cut}} \leq 0.01$. 
\item The difference between fit results in \as\ when exchanging the 
  hadronisation model is considered as an estimate of the error due to
  simulation:  
  This error already includes quark mass effects since the \apacic\ 
  model takes b-quark masses into account. 
\item To estimate the theoretical error the scale is varied around its 
  optimised value: 
  $0.5\cdot x_\mu^{\mathrm{opt}} \leq x_\mu \leq 2\cdot x_\mu^{\mathrm{opt}}$
  as in ~\cite{Abreu:2000ck}, 
  covering the scatter of experimentally optimised scales obtained with 
  different fit ranges and for different hadronisation models, see 
  \fig{fig:opt-scales}. 
\item While b-quark mass effects are included in the hadronisation corrections
  performed with APACIC++ the DEBRECEN generator used to compute the
  coefficient function in Equation \ref{eq:r4_nlo} is available only for the 
  massless case. From recent investigations of b quark mass effects
  \cite{Ballestrero:2000ur} it has been evaluated that these can shift the
  result by as much as 1.8\%. Conservatively a contribution to the uncertainty
  of this size has been  added.  

\end{itemize} 
The statistical error, the uncertainties obtained by varying track 
and event cuts and by varying \ycut\ are combined into the 
experimental error. Table \ref{tab:syserr_asmz} summarises the 
contributions to the error on the \asmz\ measurement and Table 
\ref{tab:results_as} contains the \asmz\ results. Within the 
experimental error the results obtained by using the \durham\ or 
the \cambridge\ algorithm are consistent. The total error on the 
measurement is 3.0\% for \durham\ and 2.6\% for \cambridge. If the 
scale is varied around its optimised value within the larger range 
$0.25\cdot x_\mu^{\mathrm{opt}} \leq x_\mu \leq 4\cdot 
x_\mu^{\mathrm{opt}}$ the contribution to the error on \asmz\ due 
to the $x_{\mu}$ variation has to be increased from 0.0014 to 
0.0085 for \durham\ and from 0.0007 to 0.0037 for \cambridge. 
\begin{table}[htbp] 
\begin{center} 
\renewcommand{\arraystretch}{1.2} 
\begin{tabular}{|c|r@{.}l|r@{.}l|} 
\hline 
 contribution to error       & \multicolumn{2}{|c|}{\durham} & 
                               \multicolumn{2}{|c|}{\cambridge} \\ \hline\hline 
 statistical error           & \,\, 
                               0&00045 & \,\,\,\,\, 
                                         0&00050 \\  
 cut variations              & 0&00041 & 0&00020 \\  
 working point variation     & 0&0011  & 0&0008  \\ \hline  
 total experimental error    & 0&0012  & 0&0010  \\ \hline\hline  
 MC model exchange           & 0&0023  & 0&0017  \\  
 b mass effect               & 0&0021  & 0&0021  \\ \hline 
total had. error             & 0&0031  & 0&0027  \\ \hline \hline
 \xmu\ variation             & 0&0014  & 0&0007  \\ \hline\hline 
total error on \asmz\       & 0&0036  & 0&0030  \\ 
\hline 
\end{tabular} 
\renewcommand{\arraystretch}{1.0} 
\end{center} 
\myhangcaption{xxx}{\label{tab:syserr_asmz}Contribution to the error on \asmz\ 
  for \durham\ and \cambridge.} 
\end{table} 
\begin{table}[htbp] 
\begin{center} 
\renewcommand{\arraystretch}{1.2} 
\begin{tabular}{|c|r@{.}lcr@{.}lcr@{.}lcr@{.}l|} 
\hline 
 observable   & \multicolumn{2}{|c}{\asmz} &$\pm$& 
                \multicolumn{2}{c}{exp.}   &$\pm$& 
                \multicolumn{2}{c}{hadr.}  &$\pm$& 
                \multicolumn{2}{c|}{scale}  \\ \hline \hline 
 \durham\     & 0&1178 &$\pm$& 0&0012 &$\pm$& 0&0031 &$\pm$& 0&0014 \\ \hline 
 \cambridge\  & 0&1175 &$\pm$& 0&0010 &$\pm$& 0&0027 &$\pm$& 0&0007 \\ 
\hline 
\end{tabular} 
\renewcommand{\arraystretch}{1.0} 
\end{center} 
\myhangcaption{xxx}{\label{tab:results_as}Results in \asmz\ for \durham\ and 
  \cambridge} 
\end{table} 

\clearpage 
 
\subsection{Measurement of the running of \boldmath\as \label{run}} 
%
To investigate the energy dependence of the strong coupling constant \as, the 
fits to the four-jet rates (with optimised scales, as determined 
in \sect{subsec:xmu}) are repeated at all centre-of-mass energies 
listed in \tab{tab:energien}. 
In order to account for the lower statistics of the \leptwo\ data, 
the working points are shifted to smaller values of \ycut\ which, however, are
still in the range of stable \as\ results: \durham\ 
$y_{\mathrm{cut}}= 0.0040$, \cambridge\ $y_{\mathrm{cut}}= 0.0025$. 
 
The determination of $\alpha_s$ at different energies allows the 
predicted scale dependence of the coupling due to higher 
order effects to be tested. Starting from the renormalisation group equation: 
\begin{equation} 
  \label{eq:rge} 
  E_{\mathrm {cm}}^2\frac{{\mathrm d} \alpha_s}{{\mathrm d} E_{\mathrm
  {cm}}^2}=  
  \beta(\alpha_s) = -\alpha_s^2(b_0+b_1\alpha_s + \dots) \, , 
\end{equation} 
the logarithmic energy slope is obtained: 
\begin{equation} 
  \label{eq:logslope} 
  \frac{{\mathrm d} \alpha_s^{-1}}{{\mathrm d} \log(E_{\mathrm 
  {cm}})}= - \frac{2}{\alpha_s^2} \beta(\alpha_s)  = 2b_0
  (1+\frac{b_1}{b_0}\alpha_s +  \cdots)  
  = 2b_0 \left(1+\frac{b_1}{b_0^2\log{(E_{\mathrm {cm}}^2/\Lambda^2})} +\dots 
  \right) \, , 
\end{equation} 
with $b_0=(33-2n_f)/12\pi$,  $b_1=(153-19n_f)/24\pi^2$.
In 
leading order the logarithmic derivative (\eqn{eq:logslope}) is 
independent of $\alpha_s$ and $E_{\mathrm {cm}}$ and twice the 
coefficient $b_0$ of the $\beta$ function ($2b_0=1.22$ for 
$n_f=5$). Evaluating the equation in  second order results in 
a small dependence of the derivative on $\alpha_s$. Thus 
$\Lambda_{\mathrm {QCD}}$ and an appropriate energy scale have to 
be chosen in order to calculate a single value of the logarithmic 
derivative which can be compared with the experimental result. 
Using  $E_{\mathrm {cm}}=150 \pm 60 \gev$ (the average energy of our
measurements), $\Lambda=230 $\mev 
(corresponding to $\alpha_s(M_Z)=0.118$), 
and $n_f=5$ one obtains ${\mathrm  
d}\alpha_s^{-1}/{\mathrm d} \log E_{\mathrm {cm}}=1.27 \pm 0.10$. 
 
\begin{figure}[tbhp] 
\begin{center} 
\mbox{\epsfig{file=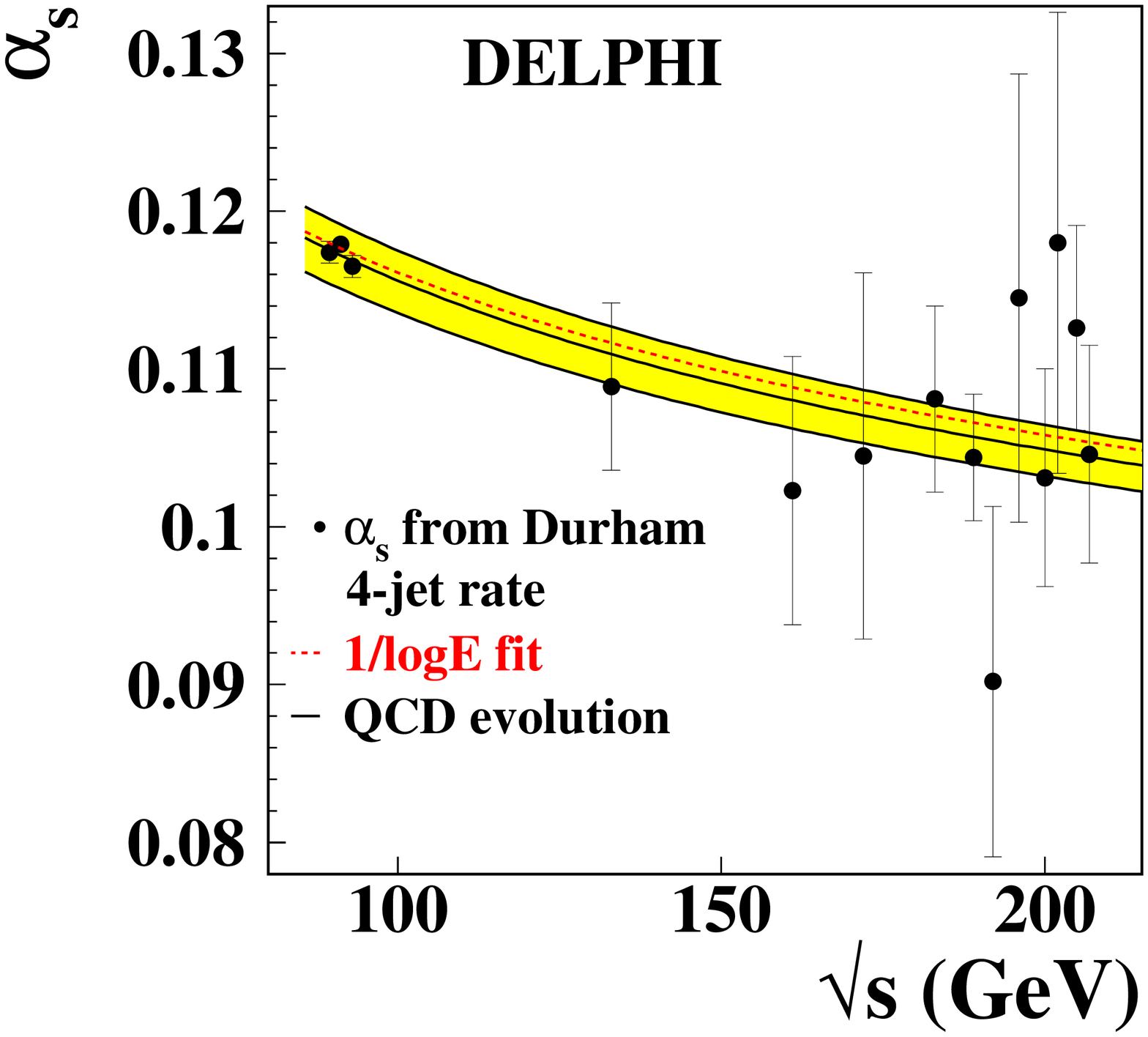,width=12.cm}} 
\mbox{\epsfig{file=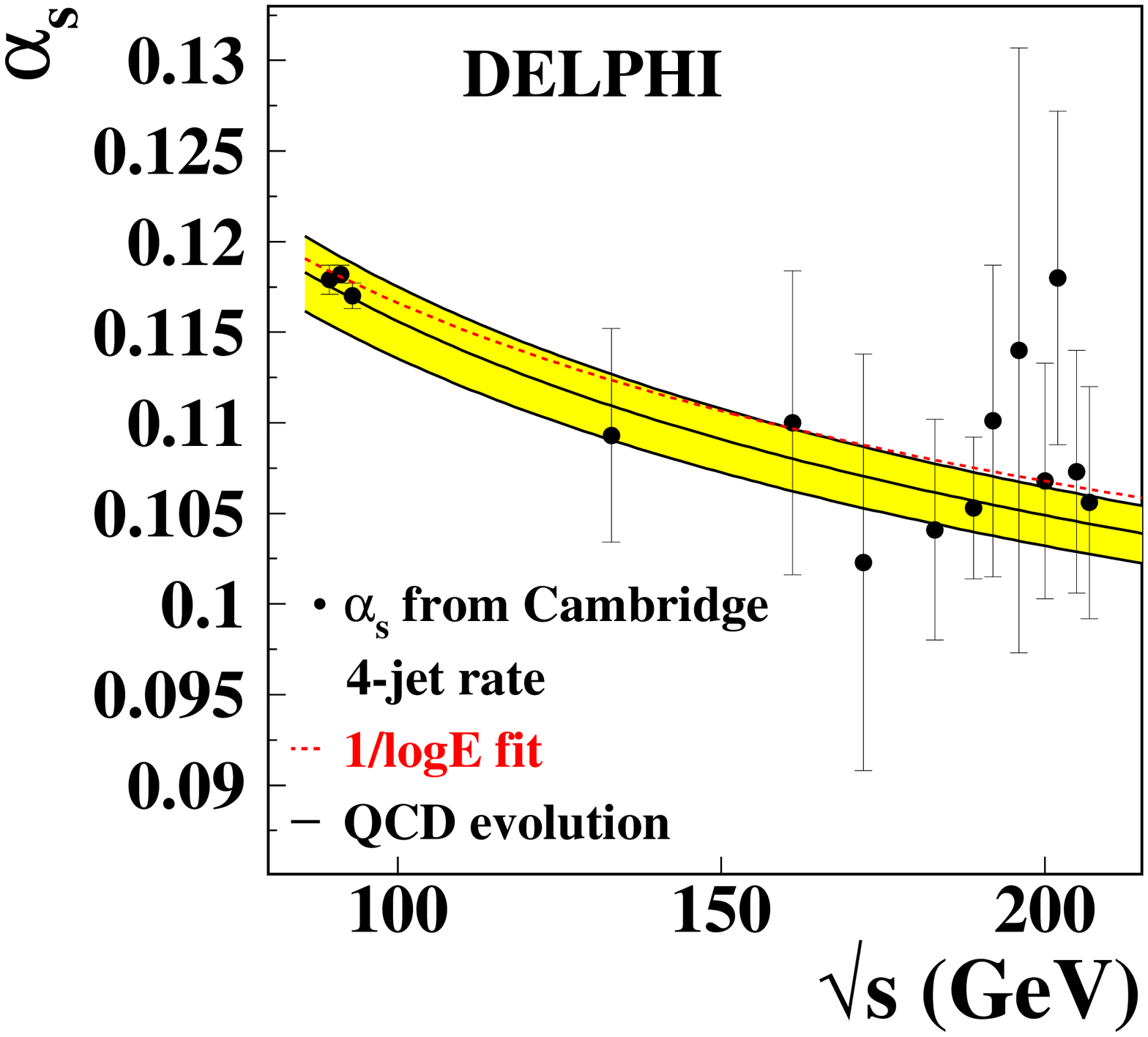,width=12.cm}} 
\end{center} 
\myhangcaption{}{\label{fig_as_mes}\label{figg_as_esd}Energy 
                     dependence of \as\ as obtained from 
                     jet rates with experimentally optimised scales. 
                     The errors shown are statistical 
                     only. The band shows the QCD 
                     expectation when extrapolating the  world average
                     \cite{Hagiwara:2002fs} to 
                     other energies. The dashed lines show the result of the
                     $1/\log\sqrt{s}$ fits. } \end{figure}

The experimental value of ${\mathrm d}\alpha_s^{-1}/{\mathrm 
d}\log{E_{\mathrm {cm}}}$ as obtained from fitting  the function 
$1/(a\log{E_{\mathrm {cm}}}+b)$ to the measured $\alpha_s$ values 
is in good agreement with the QCD expectation 
(\tab{tab_dinvasdlogE_esd} and Figure \ref{fig_as_mes}).

%
The following contributions to the systematic error on the 
logarithmic derivative are considered: 
\begin{itemize} 
\item Since the acceptance corrections $C_{acc}$ (\eqn{acc}) are 
  correlated between all energies, a possible systematic error would 
  have only a reduced influence on the logarithmic energy slope of 
  \as. Still the acceptance correction is energy-dependent. To 
  evaluate the corresponding systematic error, the difference 
  between the correction factor $C$ at the $Z$ pole and at \leptwo\ 
  energies is added to $C$ at the three energies near the $Z$ resonance at 
  89, 91, and 93 \gev\ and the fit is repeated. The full 
  difference in the slopes found with or without this 
  change is considered as the contribution to the systematic error 
  of the logarithmic slope due to the acceptance correction. 
\item At \leptwo\ energies the cut in the reconstructed centre-of-mass energy 
  is changed from  $\sqrt{s^{\prime}_{\mathrm{rec}}} \geq 0.9\cdot
  E_{\mathrm{cm}}$  
  to $\sqrt{s^{\prime}_{\mathrm{rec}}} \geq 0.8\cdot 
  E_{\mathrm{cm}}$ and  the fit is repeated. The difference in the
  logarithmic  
  slopes is taken as the contribution to the systematic error. 
\item The treatment of 4f background is an important source of systematic
  uncertainties.  
  \begin{itemize} 
    \item The 4f simulation is performed by using alternatively \pythia\ or 
    \excalibur, the full difference being included as
    the systematic error.  
    \item For the subtraction of 4f background the cross-section is varied 
      by its total error on $\pm 1.5\%$. 
    \item The cut in the discriminating variable is varied: $D^2 > 
      (900\pm100)\gev^2$. 
  \end{itemize} 
\item The renormalisation scale is varied at all energies: $1/2\cdot 
  x_\mu^{\mathrm{opt}} \leq x_\mu \leq 2\cdot x_\mu^{\mathrm{opt}}$. 
\end{itemize} 
 
Effects due to track and event selections are regarded as fully 
correlated between the energies and thus neglected. Table 
\ref{tab_dinvasdlogE_esd} contains the 
results of the ${\mathrm d}\alpha_s^{-1}/{\mathrm 
d}\log{E_{\mathrm {cm}}}$ measurements for the \durham\ and 
\cambridge\ algorithms and the corresponding statistical and 
systematical errors. 
%
%
%
\begin{table}[tbhp] 
\begin{center} 
\renewcommand{\arraystretch}{1.2} 
\begin{tabular}{|c|c|} 
\hline 
 Observable                  & ${\mathrm d}\alpha_s^{-1}/{\mathrm d}\log{E_{\mathrm {cm}}}$\\ 
\hline\hline 
 \durham\                    & $ 1.21 \pm 0.26 \pm 0.20$\\ 
 \cambridge\                 & $ 1.14 \pm 0.25 \pm 0.26$\\ 
\hline\hline 
  QCD expectation            & $1.27$\\ 
\hline 
\end{tabular} 
\end{center} 
\myhangcaption{}{\label{tab_dinvasdlogE_esd}Results 
                 of the $\mathrm{d}\alpha_s^{-1}/ \mathrm{d}\log E_{\mathrm {cm}}$ 
                 measurements for the \durham\ and \cambridge\ algorithms. The 
                 theoretical expectation is calculated in second 
                 order.} 
\end{table} 
%
 

\section{Summary} 
Hadronic jet rates in electron-positron annihilation have been 
measured by DELPHI at centre-of-mass energies between 89.4 and 209 \gev. 
The data agree with the expectation from QCD-based event generators. No 
indication of a significant excess of multijet events at high 
energies is found. 
 
The strong coupling constant has been determined from the four-jet 
rate in \oass. A variety of methods to solve the renormalisation 
scale problem has been investigated. A consistent measurement of 
\as\ can be performed by using experimentally optimised scales. 
The results obtained with two different jet-clustering algorithms 
agree. The final result quoted is obtained by applying the 
\cambridge\ algorithm, since this algorithm has small 
third-order contributions, and shows a smaller 
dependence on the renormalisation scale: 
\begin{equation} 
  \label{eq:res_as} 
  \mbox{\asmz} = 0.1175 \pm 0.0030 \, ({\mathrm {tot}}) \, . 
\end{equation} 
The result in \oass\ is statistically uncorrelated and in good 
agreement with previous \delphi\ measurements \cite{Abreu:2000ck} 
and also with the world average value \cite{Hagiwara:2002fs}. The \as\ result
is also in good  
agreement with recent \as\ measurements of the OPAL 
\cite{Abbiendi:2001qn} and ALEPH \cite{Heister:2002tq} 
collaborations based on four-jet rates measured at the Z resonance 
using \oass\ calculations combined with the resummation of large 
logarithms.  The scale-setting methods obtained in \cite{Abreu:2000ck} are 
confirmed.

The comparison of $\alpha_s$ as measured at the $Z$ and at higher 
energies confirms that the energy dependence of the strong 
coupling is consistent with QCD expectation. Results from \durham\ 
and \cambridge\ are consistent. The logarithmic energy slope, 
again obtained from \cambridge\ and again statistically 
uncorrelated to the result of the \oas\ analysis presented in 
\cite{Abreu:2003}, is measured to be 
\begin{equation} 
  \label{eq:res_runn} 
\frac{{\mathrm d}\alpha_s^{-1}}{{\mathrm d} \log E_{\mathrm {cm}}} 
         =1.14 \pm 0.36 \, ({\mathrm {tot}}) \, , 
\end{equation} 
while the QCD prediction for this quantity is 1.27. The 
measurement is in good agreement with previous measurements in 
\oas\ \cite{Abreu:1999rc,Abreu:2003}.

\subsection*{Acknowledgements} 
\vskip 3 mm 
  We are greatly indebted to our technical 
 collaborators, to the members of the CERN-SL Division for the excellent 
performance of the LEP collider, and to the funding agencies for 
their support in building and operating the DELPHI detector.\\ We 
acknowledge in particular the support of \\ Austrian Federal 
Ministry of Science and Traffics, GZ 616.364/2-III/2a/98, \\ 
FNRS--FWO, Belgium,  \\ FINEP, CNPq, CAPES, FUJB and FAPERJ, 
Brazil, \\ Czech Ministry of Industry and Trade, GA CR 202/96/0450 
and GA AVCR A1010521,\\ Danish Natural Research Council, \\ 
Commission of the European Communities (DG XII), \\ Direction des 
Sciences de la Mati$\grave{\mbox{\rm e}}$re, CEA, France, \\ 
Bundesministerium f$\ddot{\mbox{\rm u}}$r Bildung und Forschung, 
Germany,\\ General Secretariat for Research and Technology, 
Greece, \\ National Science Foundation (NWO) and Foundation for 
Research on Matter (FOM), The Netherlands, \\ Norwegian Research 
Council,  \\ State Committee for Scientific Research, Poland, 
2P03B06015, 2P03B03311 and SPUB/P03/178/98, \\ JNICT--Junta 
Nacional de Investiga\c{c}\~{a}o Cient\'{\i}fica e 
Tecnol$\acute{\mbox{\rm o}}$gica, Portugal, \\ Vedecka grantova 
agentura MS SR, Slovakia, Nr. 95/5195/134, \\ Ministry of Science 
and Technology of the Republic of Slovenia, \\ CICYT, Spain, 
AEN96--1661 and AEN96-1681,  \\ The Swedish Natural Science 
Research Council,      \\ Particle Physics and Astronomy Research 
Council, UK, \\ Department of Energy, USA, DE--FG02--94ER40817. \\ 



\end{document}